\documentclass{article}

\usepackage[english]{babel}
\usepackage[utf8]{inputenc} 
\usepackage[T1]{fontenc}    

\usepackage{arxiv}
\usepackage{booktabs}       
\usepackage{amsfonts}       
\usepackage{nicefrac}       
\usepackage{microtype}      
\usepackage{graphicx}       
\usepackage{academicons}    
\usepackage{xcolor}         
\usepackage{orcidlink}      
\usepackage{tabularx}       
\usepackage{float}          
\usepackage{natbib}
\usepackage{doi}
\usepackage{supertabular}
\usepackage{svg}
\usepackage{comment}
\usepackage{fancyvrb}
\usepackage{listings}

\usepackage{subcaption}

\usepackage{xcolor}
\lstdefinestyle{mypython}{
    language=Python,
    frame=single,
    basicstyle=\footnotesize\ttfamily,
    keywordstyle=\color{blue},
    commentstyle=\color{darkgreen},
    stringstyle=\color{red},
    showstringspaces=false,
}

\graphicspath{{./images/}}

\usepackage{tikz}
\usetikzlibrary{shapes.geometric, arrows}

\usepackage{hyperref}       

\title{\textit{vailá}: Versatile Anarcho Integrated Liberation Ánalysis in Multimodal Toolbox}

\author{
  Paulo Roberto Pereira Santiago\orcidlink{0000-0002-9460-8847}$^{1,2}$ \\
  \texttt{paulosantiago@usp.br} \\
  \And
  Abel Gonçalves Chinaglia\orcidlink{0000-0002-6955-7187}$^{2}$ \\
  \texttt{abel.chinaglia@usp.br} \\
  \And
  Kira Flanagan\orcidlink{0000-0003-0317-6346}$^{6}$ \\
  \texttt{kira.flanagan@unf.edu} \\
  \And
  Bruno L. S. Bedo\orcidlink{0000-0003-3821-2327}$^{3}$ \\
  \texttt{bruno.bedo@usp.br} \\
  \And
  Ligia Yumi Mochida\orcidlink{0009-0005-7266-3799}$^{4,5}$ \\
  \texttt{l.mochida@unf.edu} \\
  \And
  Juan Aceros\orcidlink{0000-0001-6381-7032}$^{4,6}$ \\
  \texttt{juan.aceros@unf.edu} \\
  \And
  Aline Bononi\orcidlink{0000-0001-8169-0864}$^{7}$ \\
  \texttt{bononialine@gmail.com} \\
  \And
  Guilherme Manna Cesar\orcidlink{0000-0002-5596-9439}$^{4,5}$ \\
  \texttt{g.cesar@unf.edu} \\
}

\date{
  $^1$ Biomechanics and Motor Control Laboratory, School of Physical Education and Sport, USP, Brazil \\
  $^2$ Graduate Program in Rehabilitation and Functional Performance, Medical School, USP, Brazil \\
  $^3$ Technology and Sports Performance Analysis Laboratory, School of Physical Education, USP, Brazil \\
  $^4$ Laboratory of Applied Biomechanics and Engineering, Brooks College of Health, UNF, USA \\
  $^5$ Department of Physical Therapy, Brooks College of Health, UNF, USA \\
  $^6$ College of Computing, Engineering and Construction, UNF, USA \\
  $^7$ Municipal Pharmacy of Ribeirão Preto, Brazil 
}

\hypersetup{
pdftitle={\textit{vailá}: Versatile Anarcho Integrated Liberation Ánalysis in Multimodal Toolbox},
pdfsubject={biomechanics, multimodal analysis, open-source},
pdfauthor={Paulo Roberto Pereira Santiago, Abel Gonçalves Chinaglia, Bruno Luiz de Souza Bedo, Ligia Yumi Mochida, Juan Aceros, Guilherme Manna Cesar},
pdfkeywords={biomechanics, human movement analysis, open-source software, multimodal data integration},
 colorlinks     = true,
  linkcolor      = black,            
  citecolor      = black,            
  filecolor      = black,            
  urlcolor       = blue,
  bookmarksdepth = 4}

\begin{document}
\maketitle

\begin{abstract}
Human movement analysis is crucial in health and sports biomechanics for understanding physical performance, guiding rehabilitation, and preventing injuries. However, existing tools are often proprietary, expensive, and function as "black boxes", limiting user control and customization. This paper introduces \textit{vailá}—Versatile Anarcho Integrated Liberation Ánalysis in Multimodal Toolbox—an open-source, Python-based platform designed to enhance human movement analysis by integrating data from multiple biomechanical systems. \textit{vailá} supports data from diverse sources, including retroreflective motion capture systems, inertial measurement units (IMUs), markerless video capture technology, electromyography (EMG), force plates, and GPS/GNSS systems, enabling comprehensive analysis of movement patterns. Developed entirely in Python 3.11.9, which offers improved efficiency and long-term support, and featuring a straightforward installation process, \textit{vailá} is accessible to users without extensive programming experience. In this paper, we also present several workflow examples that demonstrate how \textit{vailá} allows the rapid processing of large batches of data, independent of the type of collection method. This flexibility is especially valuable in research scenarios where unexpected data collection challenges arise, ensuring no valuable data point is lost. We demonstrate the application of \textit{vailá} in analyzing sit-to-stand movements in pediatric disability, showcasing its capability to provide deeper insights even with unexpected  movement patterns. By fostering a collaborative and open environment, \textit{vailá} encourages users to innovate, customize, and freely explore their analysis needs, potentially contributing to the advancement of rehabilitation strategies and performance optimization.
For more details in GitHub repository: \\
\url{https://github.com/vaila-multimodaltoolbox/vaila} and the online documentation: \url{https://vaila.readthedocs.io}.
\end{abstract}


\section{Introduction}

Human movement analysis is essential for the understanding of physical performance, rehabilitation, and injury prevention within health and sports biomechanics~\citep{ricamato2005quantification,roberts2017,cesar2016frontal,bedo2021influence,cesar2022muscle}. Accurate analysis of human movement patterns can lead to improved rehabilitation strategies, enhanced athletic performance, and a deeper understanding of musculoskeletal function. However, existing tools for movement analysis often present significant limitations. For instance, the Biomechanical Tool Kit (BTK) Mokka, once a popular biomechanical analysis choice, has not been updated or developed for over nine years~\citep{Barre2014}. Many contemporary solutions are proprietary, expensive, and function as "black boxes," restricting user control and customization and hindering the advancement of research through community contributions.

Several prominent tools in the field, such as OpenSim, are widely used for musculoskeletal modeling and simulation~\citep{seth2018opensim}. OpenSim allows for the development of custom toolboxes, which has been instrumental in biomechanics research. For example, custom musculoskeletal models have been developed to estimate tibiofemoral contact forces during tasks involving high knee and hip flexion~\citep{bedo2020custom}. Similarly, BOPS, a MATLAB toolbox, facilitates batch processing of musculoskeletal data in OpenSim~\citep{bedo2020474}. Despite their utility, these tools rely heavily on programming languages like C++ and MATLAB, the latter requiring a paid license. This reliance can pose barriers for researchers without access to such software, limiting the accessibility and dissemination of advanced biomechanical analysis methods.

In contrast, Python is a high-level, open-source programming language known for its simplicity, extensive libraries, and strong community support. Python offers enhanced memory safety and portability across platforms such as Linux, macOS, and Windows~\citep{python311}. Leveraging Python's capabilities, we have developed \textit{vailá}—\textit{Versatile Anarcho Integrated Liberation Ánalysis in Multimodal Toolbox}—an open-source, customizable platform designed to enhance human movement analysis by integrating data from multiple biomechanical systems. \textit{vailá} supports data from diverse sources, including retroreflective motion capture systems, inertial measurement units (IMUs), markerless video capture technology, electromyography (EMG), force plates, and GPS/GNSS systems, enabling comprehensive and precise analysis of movement patterns.

The motivation for developing \textit{vailá} arose from the need for a flexible, open-source tool capable of facilitating batch processing of large datasets in biomechanical research environments. During our collaboration with the University of North Florida’s Laboratory of Applied Biomechanics and Engineering (LABE) under the CAPES-PAME program, we identified data processing workflows as a bottleneck in research activities. The increasing use of markerless solutions for participants with disabilities or comprehension difficulties highlighted the necessity for a tool that could simplify batch processing, integrate multimodal data, and provide researchers with the flexibility to modify the software according to their needs.

The main objective of this paper is to present and outline the development of \textit{vailá}, which was initially designed to serve as a multimodal processing tool in biomechanical data collection for children with cerebral palsy. \textit{vailá} acts as a facilitator, inviting individuals without extensive expertise in programming or biomechanics to contribute and enhance the software. By providing an open, modular, and inclusive platform for data analysis, \textit{vailá} empowers the biomechanics research community to innovate, customize, and freely explore their analysis needs, thereby breaking down barriers imposed by traditional software.

\section{Related Work}

Several existing tools and frameworks have significantly contributed to the field of biomechanics and multimodal data analysis. OpenSim is one of the most established platforms for musculoskeletal modeling and simulation~\citep{seth2018opensim}. It allows users to extend its functionalities through custom toolboxes, which has been instrumental in advancing biomechanics research. For instance, custom musculoskeletal models have been developed using OpenSim to estimate tibiofemoral contact forces during activities involving high knee and hip flexion~\citep{bedo2020custom}. BOPS (Batch OpenSim Processing Scripts), a MATLAB toolbox, was also created to facilitate batch processing of musculoskeletal data~\citep{bedo2020474}. However, the reliance on MATLAB, a proprietary platform, introduces financial and accessibility constraints, potentially limiting the tool's adoption in resource-limited settings.

Other solutions, such as BTK Mokka, provided a free and open-source alternative for biomechanics research~\citep{Barre2014}. Despite its initial utility, BTK Mokka has seen no active development for several years, leading to gaps in functionality and compatibility with modern technologies. Its limitations include lacking support for recent hardware advancements and data formats, which hinders its adoption in contemporary research environments.

Markerless motion capture has gained traction due to its potential to streamline data collection, especially for participants who may not tolerate traditional markers~\citep{Smith2023}. Platforms like MediaPipe offer pipelines for human pose estimation using machine learning techniques~\citep{lugaresi2019mediapipe}. While MediaPipe is a powerful tool for general pose estimation, it is not specifically tailored for biomechanics research, and its integration with biomechanical analysis pipelines remains underdeveloped.

In the realm of open-source tools, DeepLabCut~\citep{Matthis2022} has made significant strides by providing a user-friendly platform for markerless pose estimation across species and behaviors. However, it primarily focuses on pose estimation and does not offer comprehensive biomechanical analysis or multimodal data integration.

Similarly, free tools like Tracker~\citep{wee2013open}, Kinovea, and Dvideow~\citep{figueroa2003flexible} have been instrumental in advancing biomechanical analyses by providing accessible platforms for motion tracking and analysis. While these tools have significantly contributed to the field, they cannot often handle multiple data modalities or require extensive manual processing, which can be time-consuming and less efficient for large datasets.

In contrast, \textit{vailá} offers a flexible solution that integrates multiple data sources, such as motion capture, IMU, EMG, and supports markerless data collection. Developed in Python 3.11.9—which benefits from CPython acceleration—\textit{vailá} leverages the extensive libraries and strong community support inherent to Python. This choice of programming language enhances performance and makes the tool more accessible to researchers, as Python is widely taught and used across scientific disciplines.

By providing an open-source, Python-based platform, \textit{vailá} lowers the entry barrier for biomechanics researchers and facilitates community-driven development. This approach allows for rapid integration of new data sources and technologies, ensuring the toolbox stays relevant in a rapidly evolving research landscape. The transparency of the codebase encourages users to engage with and modify the software, fostering an environment where imagination and innovation drive the advancement of biomechanical analysis.

\section{Methods}

\subsection{Software Engineering Principles in \textit{vailá}}

The design and development of the \textit{vailá} toolbox are rooted in well-established software engineering principles, particularly those outlined by Frederick P. Brooks in \textit{The Mythical Man-Month}~\citep{brooks1995mythical, brooks1995mythical_20years}. \textit{vailá} was conceived as a modular, open-source, and transparent system to meet the needs of a biomechanics and bioengineering research lab in its early stages of operation.

In a research environment, especially in biomechanics, planned methods and equipment cannot always be applied as intended due to factors such as participants' disabilities or limited comprehension. These circumstances often require adjustments to data collection methods. To accommodate this variability, \textit{vailá} was designed with flexibility in mind, allowing for easy modifications to the software.

One of the primary goals of \textit{vailá} is to ensure that the system remains open, modular, and extensible, enabling users and contributors to modify or add new features easily. Every Python script in \textit{vailá} includes mechanisms for transparency, where the code prints essential metadata—such as the name and path of the script being executed—providing users with a clear understanding of what is running and where. Additionally, whenever possible, \textit{vailá} ensures that all important calculations and relevant procedures are displayed in the terminal, giving users full visibility into the ongoing processes.

For example, each script in \textit{vailá} uses the following print statements to provide transparency about its execution:

\begin{verbatim}
import os
from rich import print

# Print the directory and name of the script being executed
print(f"Running script: {os.path.basename(__file__)}")
print(f"Script directory: {os.path.dirname(os.path.abspath(__file__))}")
\end{verbatim}

This transparency enables users to track the execution flow, understand key calculations, debug more efficiently, and better comprehend the modular structure of the software. Using the \texttt{rich} library enhances terminal output, making it easier to visualize the steps in the analysis process.

By adhering to these principles, \textit{vailá} allows researchers to adapt the software to their specific needs while maintaining clarity about what is being executed. This simplifies debugging and the development of new features and ensures that users are informed about the key operations and calculations being performed throughout the analysis.

\subsection{Modular Architecture of \textit{vailá} GUI and CLI}

The core design philosophy of \textit{vailá} is its modularity. Each major toolbox functionality is encapsulated in self-contained modules, allowing individual components to be developed, tested, and deployed independently. This approach follows the principle of separation of concerns, where each module is responsible for a specific task, such as file management, data analysis, or visualization.

The graphical user interface (GUI) of \textit{vailá} is divided into three main frames: File Manager (Frame A), Multimodal Analysis (Frame B), and Available Tools (Frame C). Each frame is organized into rows and columns, where each button represents a specific functionality related to file handling, multimodal data analysis, or available tools for further processing and visualization. Below is the detailed explanation of each frame:

\begin{itemize}
    \item \textbf{Frame A: File Manager} – Responsible for file operations, this section provides buttons for managing files and directories. The rows and columns are structured as follows:
    \begin{itemize}
        \item \textbf{A\_r1\_c1 to A\_r1\_c9:} 
        \begin{itemize}
            \item A\_r1\_c1 – Rename Files
            \item A\_r1\_c2 – Import Files
            \item A\_r1\_c3 – Export Files
            \item A\_r1\_c4 – Copy Files
            \item A\_r1\_c5 – Move Files
            \item A\_r1\_c6 – Remove Files
            \item A\_r1\_c7 – Tree View
            \item A\_r1\_c8 – Find Files
            \item A\_r1\_c9 – Transfer Files
        \end{itemize}
    \end{itemize}

    \item \textbf{Frame B: Multimodal Analysis} – This frame includes tools for analyzing various types of biomechanical data such as IMU, motion capture (MoCap), and markerless video analysis. The rows and columns are organized as follows:
    \begin{itemize}
        \item \textbf{B1\_r1\_c1 to B1\_r1\_c5:} 
        \begin{itemize}
            \item B1\_r1\_c1 – IMU Analysis
            \item B1\_r1\_c2 – MoCap Cluster Analysis
            \item B1\_r1\_c3 – MoCap Full Body Analysis
            \item B1\_r1\_c4 – Markerless 2D Analysis
            \item B1\_r1\_c5 – Markerless 3D Analysis
        \end{itemize}
        \item \textbf{B2\_r2\_c1 to B2\_r2\_c5:}
        \begin{itemize}
            \item B2\_r2\_c1 – Vector Coding
            \item B2\_r2\_c2 – EMG Analysis
            \item B2\_r2\_c3 – Force Plate Analysis
            \item B2\_r2\_c4 – GNSS/GPS Data
            \item B2\_r2\_c5 – MEG/EEG Data
        \end{itemize}
        \item \textbf{B3\_r3\_c1 to B3\_r3\_c5:}
        \begin{itemize}
            \item B3\_r3\_c1 – HR/ECG Data
            \item B3\_r3\_c2 to B3\_r3\_c5 – Placeholder for other custom tools (\textit{vailá})
        \end{itemize}
    \end{itemize}

    \item \textbf{Frame C: Available Tools} – This frame includes tools for data conversion, video and image processing, and visualization. It is organized into three sections (Data Files, Video and Image, and Visualization) and structured as follows:
    \begin{itemize}
        \item \textbf{C\_A: Data Files}
        \begin{itemize}
            \item C\_A\_r1\_c1 – Edit CSV
            \item C\_A\_r1\_c2 – C3D <--> CSV Conversion
            \item C\_A\_r1\_c3 – Placeholder for \textit{vailá} tools
            \item C\_A\_r2\_c1 to C\_A\_r3\_c3 – DLT Methods (2D and 3D) and additional placeholder tools (\textit{vailá})
        \end{itemize}
        \item \textbf{C\_B: Video and Image}
        \begin{itemize}
            \item C\_B\_r1\_c1 – Convert Video to PNG
            \item C\_B\_r1\_c2 – Cut Videos
            \item C\_B\_r1\_c3 – Draw Box
            \item C\_B\_r2\_c1 – Compress Videos (H264)
            \item C\_B\_r2\_c2 – Compress Videos (H265)
            \item C\_B\_r2\_c3 – Make Sync File
            \item C\_B\_r3\_c1 – Get Pixel Coordinates
            \item C\_B\_r3\_c2 – Metadata Information
            \item C\_B\_r3\_c3 – Merge Videos
        \end{itemize}
        \item \textbf{C\_C: Visualization}
        \begin{itemize}
            \item C\_C\_r1\_c1 – Show C3D
            \item C\_C\_r1\_c2 – Show CSV
            \item C\_C\_r2\_c1 – Plot 2D
            \item C\_C\_r2\_c2 – Plot 3D
            \item C\_C\_r3\_c1 to C\_C\_r4\_c3 – Placeholder for additional \textit{vailá} visualization tools
        \end{itemize}
    \end{itemize}
\end{itemize}

Each frame (A, B, C) is designed to handle a distinct part of the data processing and analysis workflow, allowing the user to move through data management, multimodal analysis, and tools for visualization or file conversion seamlessly. The modular layout also enables future expansion and customization by adding new tools or features in placeholders indicated as \textit{vailá}. This structure provides flexibility and ease of use, tailored for biomechanical data processing needs.

Each area is further divided into individual components or scripts, which can be customized or expanded upon. The diagram emphasizes the modularity of the \textit{vailá} toolbox. The process starts with file management, progresses through multimodal data processing and analysis, and concludes with generating results, which can be visualized and exported in various formats.

\begin{figure}[htbp]
    \centering
    \includegraphics[width=0.8\textwidth]{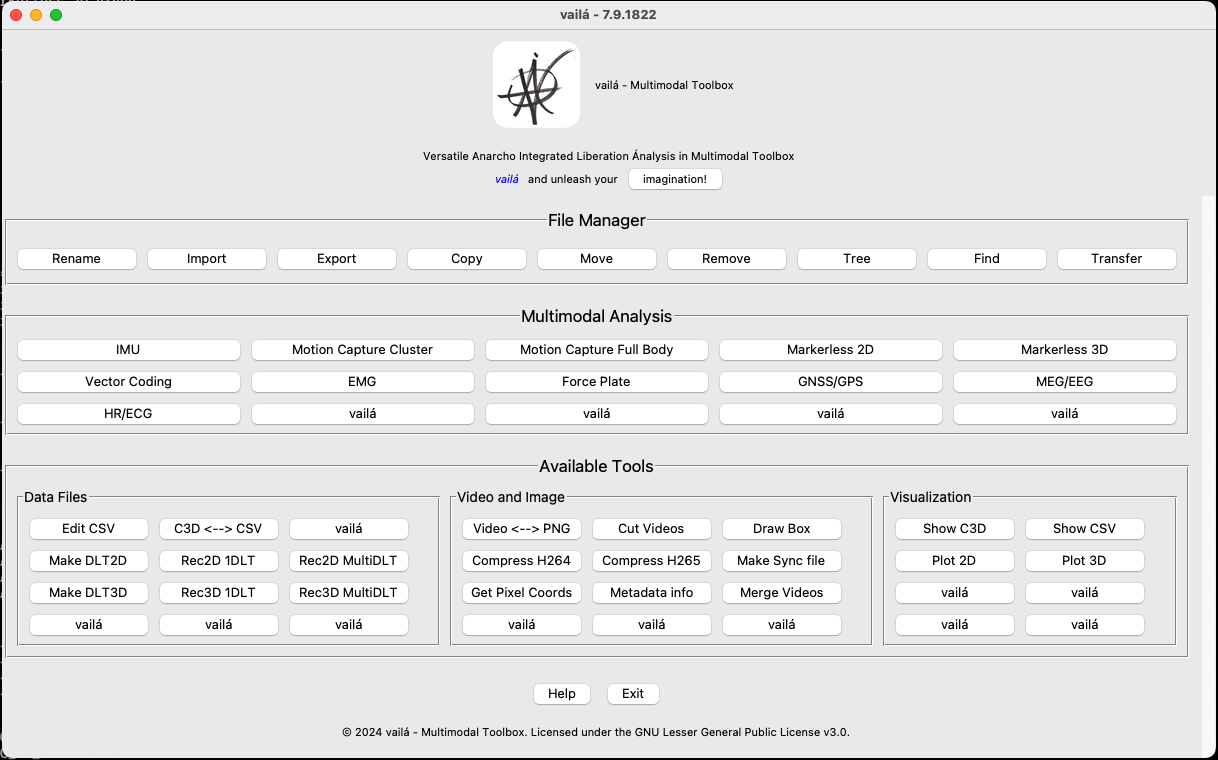}
    \caption{Graphical User Interface (GUI) of the \textit{vailá} toolbox, built using the \textit{Tkinter} library in Python. The interface is organized into three hierarchical levels. The first level consists of three main frames: \textbf{File Manager} (Frame A) for file operations, \textbf{Multimodal Analysis} (Frame B) for handling biomechanical data such as MoCap, IMU, and GNSS, and \textbf{Available Tools} (Frame C) for additional functionalities like visualization and file conversion. The second and third levels within each frame represent the rows and columns, where specific tools and operations are accessible. This modular structure allows for seamless data processing and analysis while providing placeholders labeled "vailá" for future tool expansions and customizations, ensuring flexibility and ease of use.}
    \label{fig:multimodal_integration}
\end{figure}

These capabilities enable researchers to explore new dimensions in biomechanical studies by correlating data from different sources. The synchronization functionality was effective across different systems, providing accurate temporal alignment with minimal preprocessing (Figure~\ref{fig:multimodal_integration}).

\vspace{-1em}

\subsection{Transparent Code Execution with Print Statements}

Following the principle of making code execution transparent, every script in \textit{vailá} includes print statements that display the script's name and path. This approach ensures that users have complete visibility into the behind-the-scenes processes, contributing to a more intuitive understanding of the analysis workflow.

The print statements, integrated with the \texttt{rich} library for enhanced terminal output, provide the following benefits:
\begin{itemize}
  \item \textbf{Track Execution Flow:} By showing the execution flow directly in the terminal, users can follow which scripts and functions are being called during their analysis.
  \item \textbf{Efficient Debugging:} When an issue arises, the printed file paths and script names help users pinpoint where specific operations are occurring, streamlining the debugging process.
  \item \textbf{Understand Modularity:} As \textit{vailá} is designed with a modular architecture, printing the execution path allows users to see how different components interact without diving deep into the codebase.
\end{itemize}

This transparency aligns with the ethos of open-source software, empowering users to understand and control their analysis environment. The combination of print statements and structured logging ensures that users of \textit{vailá} can monitor the progress of their analyses in real-time, which is particularly important for large-scale biomechanical studies involving batch processing.

By implementing such features, \textit{vailá} not only simplifies the analysis process but also reinforces the concept of open-source collaboration, where code transparency and traceability are fundamental.

\subsection{Terminal and Debugging Enhancements}

Several tools have been integrated into the environment to streamline the development process and debugging of new modules within the \textit{vailá} toolbox. One key aspect is the installation of \href{https://xon.sh/}{Xonsh} and the use of the ipdb debugger, combined with rich terminal output.

Xonsh is a Python-powered shell designed to enhance the traditional terminal experience. It allows seamless mixing of Python commands and shell primitives, combining the functionalities of Bash and Python in one environment. This provides a significant advantage for developers and users who need powerful scripting and shell interaction.

Xonsh is a cross-platform shell that works on Linux, macOS, and Windows, making it a suitable choice for the \textit{vailá} toolbox, which targets users from different operating systems. The command-line interface can be accessed by invoking Xonsh in the terminal or by clicking the "imagination!" button in the \textit{vailá} graphical user interface (GUI), which opens the Xonsh terminal within the active Conda environment.

\subsubsection{Python Debugging with \texttt{ipdb}}

To further support the debugging process, \textit{vailá} incorporates the \texttt{ipdb} library, which provides an enhanced version of Python's standard debugger (pdb). \texttt{ipdb} includes features such as tab completion, syntax highlighting, and advanced tracebacks, improving developer productivity when tracing errors or analyzing the flow of the program. This debugging tool is available within any toolbox module and can be triggered during code execution for real-time variables and control flow inspection.

\begin{verbatim}
import ipdb
ipdb.set_trace()  # This sets a breakpoint in the code
\end{verbatim}

\subsubsection{Terminal Output with \texttt{rich}}

Additionally, the terminal output in \textit{vailá} is enhanced by the \texttt{rich} library. \texttt{rich} improves the visual output of the terminal by enabling syntax highlighting, pretty printing, and better readability of complex data structures. Every Python script executed within \textit{vailá} prints its status using \texttt{rich}, making the terminal output clearer and more user-friendly.

\begin{verbatim}
from rich import print
print("This is an enhanced output with [bold magenta]Rich[/bold magenta]!")
\end{verbatim}

\begin{figure}[htbp]
    \centering
    \includegraphics[width=0.8\textwidth]{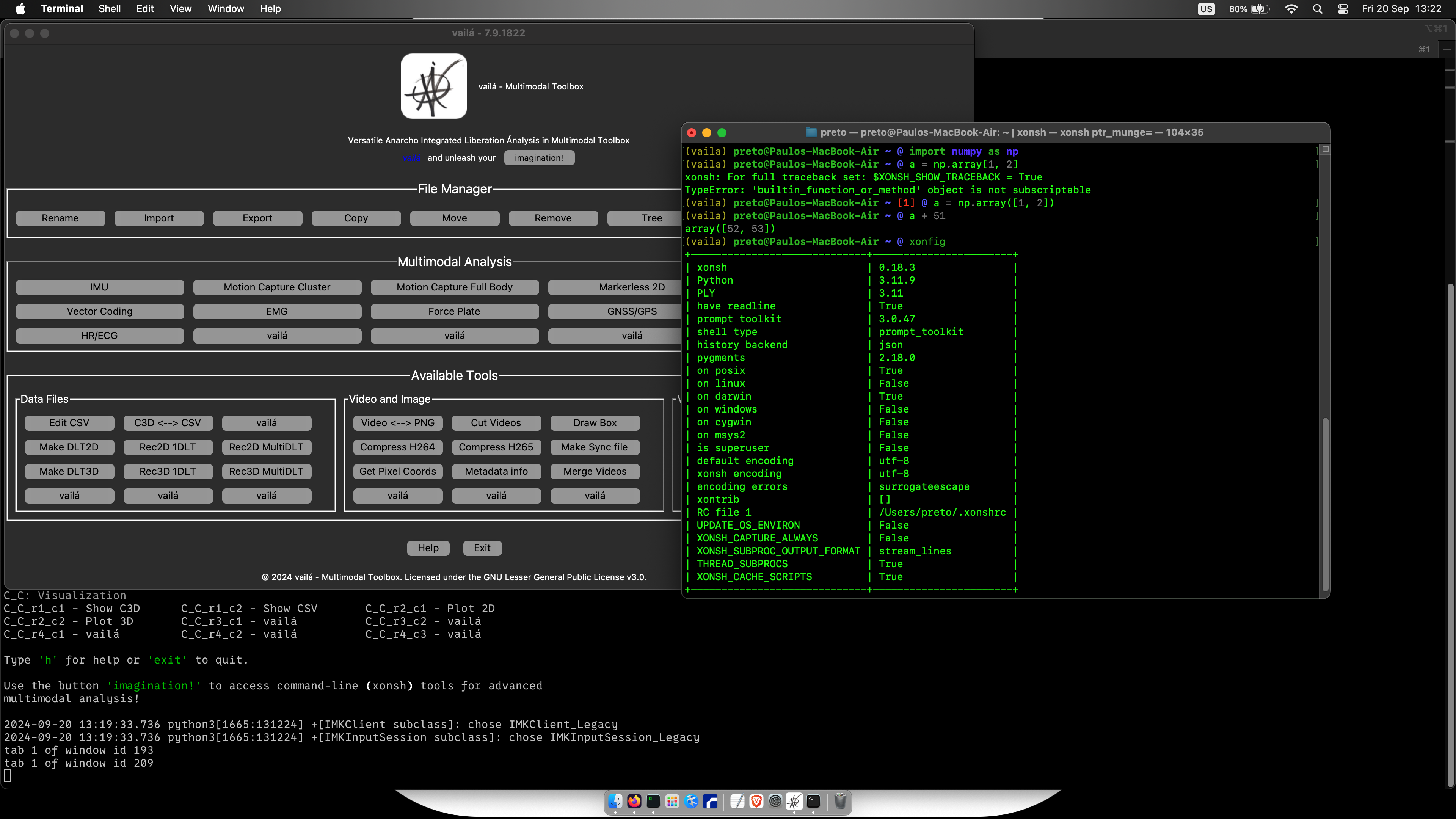}
    \caption{A terminal session running in the \textbf{Xonsh} shell, displaying enhanced output with \texttt{rich}, and real-time debugging with \texttt{ipdb}. This setup allows users to mix Python 3.11.9 and shell commands while debugging and viewing formatted terminal output.}
    \label{fig:xonsh_integration}
\end{figure}

The Figure~\ref{fig:xonsh_integration} demonstrates the enhanced terminal setup with Xonsh, \texttt{ipdb}, and \texttt{rich}, showcasing how Python commands, shell operations, and debugging coexist in the same environment.

\vspace{-0.5em}

The combination of Xonsh, \texttt{ipdb}, and \texttt{rich} makes the development and troubleshooting processes in \textit{vailá} more efficient while also providing a flexible interface for users to interact with both the shell and Python code.

As an open-source project, \textit{vailá} is licensed under the GNU Lesser General Public License version 3.0 \citep{GPLv3}. This license allows users to freely use, modify, and distribute the software if modifications to the core library are made available under the same license. This ensures that the community benefits from continuous improvements to the toolbox, fostering a collaborative development environment.

\subsection{Code Header and Documentation}

To maintain code organization and facilitate cooperation between developers, all Python scripts in the \textit{vailá} toolbox include a standardized header. This header provides essential information, such as the script's author, version, description, and instructions for usage. Additionally, each line of the code is thoroughly commented for clarity.

Below is an example of a script header, formatted using the \texttt{verbatim} environment to showcase how documentation is included in each Python script:

\scriptsize
\begin{verbatim}
"""
Script: markerless_2D_analysis.py
Author: Prof. Dr. Paulo Santiago
Version: 0.2.0
Last Updated: September 28, 2024

Description:
    This script performs batch processing of videos for 2D pose estimation using 
    MediaPipe's Pose model. It processes videos from a specified input directory, 
    overlays pose landmarks on each video frame, and exports both normalized and 
    pixel-based landmark coordinates to CSV files. 

    The user can configure key MediaPipe parameters via a graphical interface, 
    including detection confidence, tracking confidence, model complexity, and 
    whether to enable segmentation and smooth segmentation. The default settings 
    prioritize the highest detection accuracy and tracking precision, which may 
    increase computational cost.

New Features:
    - Default values for MediaPipe parameters are set to maximize detection and 
      tracking accuracy:
        - `min_detection_confidence=1.0`
        - `min_tracking_confidence=1.0`
        - `model_complexity=2` (maximum complexity)
        - `enable_segmentation=True` (segmentation activated)
        - `smooth_segmentation=True` (smooth segmentation enabled)
    - User input dialog allows fine-tuning these values if desired.

Usage:
    - Run the script to open a graphical interface for selecting the input directory 
      containing video files (.mp4, .avi, .mov), the output directory, and for 
      specifying the MediaPipe configuration parameters.
    - The script processes each video, generating an output video with overlaid pose 
      landmarks, and CSV files containing both normalized and pixel-based landmark 
      coordinates.

How to Execute:
    1. Ensure you have all dependencies installed:
       - Install OpenCV: `pip install opencv-python`
       - Install MediaPipe: `pip install mediapipe`
       - Tkinter is usually bundled with Python installations.
    2. Open a terminal and navigate to the directory where `markerless_2D_analysis.py` is located.
    3. Run the script using Python:
       
       python markerless_2D_analysis.py
       
    4. Follow the graphical interface prompts:
       - Select the input directory with videos (.mp4, .avi, .mov).
       - Select the base output directory for processed videos and CSVs.
       - Configure the MediaPipe parameters (or leave them as default for maximum accuracy).
    5. The script will process the videos and save the outputs in the specified output directory.

Requirements:
    - Python 3.11.9
    - OpenCV (`pip install opencv-python`)
    - MediaPipe (`pip install mediapipe`)
    - Tkinter (usually included with Python installations)
    - Pillow (if using image manipulation: `pip install Pillow`)

Output:
    The following files are generated for each processed video:
    1. Processed Video (`*_mp.mp4`): 
       The video with the 2D pose landmarks overlaid on the original frames.
    2. Normalized Landmark CSV (`*_mp_norm.csv`):
       A CSV file containing the landmark coordinates normalized to a scale between 0 and 1 
       for each frame. These coordinates represent the relative positions of landmarks in the video.
    3. Pixel Landmark CSV (`*_mp_pixel.csv`):
       A CSV file containing the landmark coordinates in pixel format. The x and y coordinates 
       are scaled to the video’s resolution, representing the exact pixel positions of the landmarks.
    4. Log File (`log_info.txt`):
       A log file containing video metadata and processing information, such as resolution, frame rate, 
       total number of frames, codec used, and the MediaPipe Pose configuration used in the processing.

Example:
    1. Select a folder with videos in .mp4 format.
    2. Choose the output directory for saving processed videos and CSVs.
    3. Enter the desired values for detection confidence (e.g., 0.5), tracking confidence (e.g., 0.5), 
       model complexity (0, 1, or 2), and segmentation options.
    4. The processed files will be saved with landmarks overlaid and CSVs in the chosen 
       output directory.

License:
    This program is free software: you can redistribute it and/or modify it under the terms of 
    the GNU General Public License as published by the Free Software Foundation, either version 3 
    of the License, or (at your option) any later version.

    This program is distributed in the hope that it will be useful, but WITHOUT ANY WARRANTY; 
    without even the implied warranty of MERCHANTABILITY or FITNESS FOR A PARTICULAR PURPOSE. 
    See the GNU General Public License for more details.

    You should have received a copy of the GNU GPLv3 (General Public License Version 3) along with this program. 
    If not, see <https://www.gnu.org/licenses/>.
"""
\end{verbatim}
\normalsize

This standardized documentation style ensures that each script's purpose, usage, and dependencies are clearly defined, promoting consistency across the \textit{vailá} toolbox. Moreover, including extensive comments within the codebase aids in maintaining the transparency and modularity of the project, enabling both novice and experienced developers to contribute effectively.

\subsection{Batch Processing and Automation}

One of the fundamental features of \textit{vailá} is its ability to run batch processes across multiple files, significantly improving efficiency in data analysis workflows. Instead of manually selecting and processing files one by one, \textit{vailá} automatically applies the same operations to all files in a specified directory, reducing the time needed for large-scale biomechanical studies.

This functionality is particularly advantageous for researchers in biomechanics and bioengineering laboratories, where data collection occurs frequently, often yielding large volumes of files to be processed weekly. Handling these datasets manually can be time-consuming and prone to error. Using batch processing, \textit{vailá} simplifies this process, ensuring that each file is processed similarly, with minimal user intervention. 

When files are organized appropriately in the \textbf{File Manager} (Frame A), all processing tasks can be executed efficiently in "batch" mode. This method allows researchers to automate repetitive tasks, freeing up time for higher-level analysis and decision-making. The structure of \textit{vailá} ensures that all Python scripts are designed to handle batch processing, operating in a \texttt{for} loop to process files of the same extension and internal organization.

For example, if a lab collects force platform data in CSV format or video files for motion capture analysis, \textit{vailá} can be configured to automatically process each of these files in sequence, applying the same filters, metrics calculations, or visualizations across the entire dataset. This accelerates the workflow and minimizes the risk of human error that can arise when handling large datasets manually.

The batch processing capabilities of \textit{vailá} are an essential feature for labs that deal with high volumes of data, enabling researchers to process their data in a fraction of the time it would take using manual methods. By automating these tasks, the toolbox enhances productivity, allowing researchers to focus on analyzing their results rather than managing file operations. The modular architecture ensures that batch operations are accessible across all tools, providing a streamlined and efficient workflow for processing biomechanical data.

\section{Results}

This section presents the performance results of the \textit{vailá} toolbox when processing a complete biomechanical data collection workflow. All examples provided here can be found in the \texttt{tests/} directory of the \textit{vailá} repository, which includes sample data and code to replicate the results shown. This workflow encompasses multiple types of biomechanical data (e.g., C3D, EMG, IMU, markerless 2D, and motion capture), demonstrating the flexibility and robustness of \textit{vailá}.

\subsection{Codebase Overview}

To illustrate the current state of the \textit{vailá} toolbox, we present an overview of its codebase. The project is predominantly written in Python 3.11.9, with contributions from Shell and PowerShell scripts, reflecting the diversity of tools used for automation and data processing (Table~\ref{tab:code_stats}).

\begin{table}[htbp]
    \centering
    \caption{Current codebase statistics of \textit{vailá} as shown on GitHub.}
    \label{tab:code_stats}
    \begin{tabular}{lrr}
        \toprule
        \textbf{Language} & \textbf{Percentage} & \textbf{Lines of Code} \\
        \midrule
        Python      & 94.8\%  & 15,630 \\
        Shell       & 2.7\%   & 445   \\
        PowerShell  & 2.5\%   & 256   \\
        \bottomrule
    \end{tabular}
\end{table}

Table~\ref{tab:code_stats}, the \textit{vailá} project is composed mainly of Python code (94.8\%), with Shell (2.7\%) and PowerShell (2.5

\subsection{Tracking Marker Points in Videos: The \textit{Get Pixel Coordinate} Tool}

One of the critical tools within \textit{vailá} is the \textit{Get Pixel Coordinate} tool, implemented in the script \texttt{getpixelvideo.py}. This tool allows users to manually mark and save pixel coordinates in video frames, featuring zoom functionality for precise annotations. It is particularly useful when working with markerless video data, where manual annotation or verification of automatically detected key points is necessary.

The tool supports various video formats, such as \texttt{.mp4}, \texttt{.avi}, \texttt{.mov}, and \texttt{.mkv}, and allows users to load pre-marked points from CSV files. It controls frame navigation, zooming, and marking or adjusting points. Marked coordinates are saved in CSV format, facilitating subsequent biomechanical analysis within the \textit{vailá} workflow.

\begin{figure}[htbp]
\centering
\includegraphics[width=0.8\textwidth]{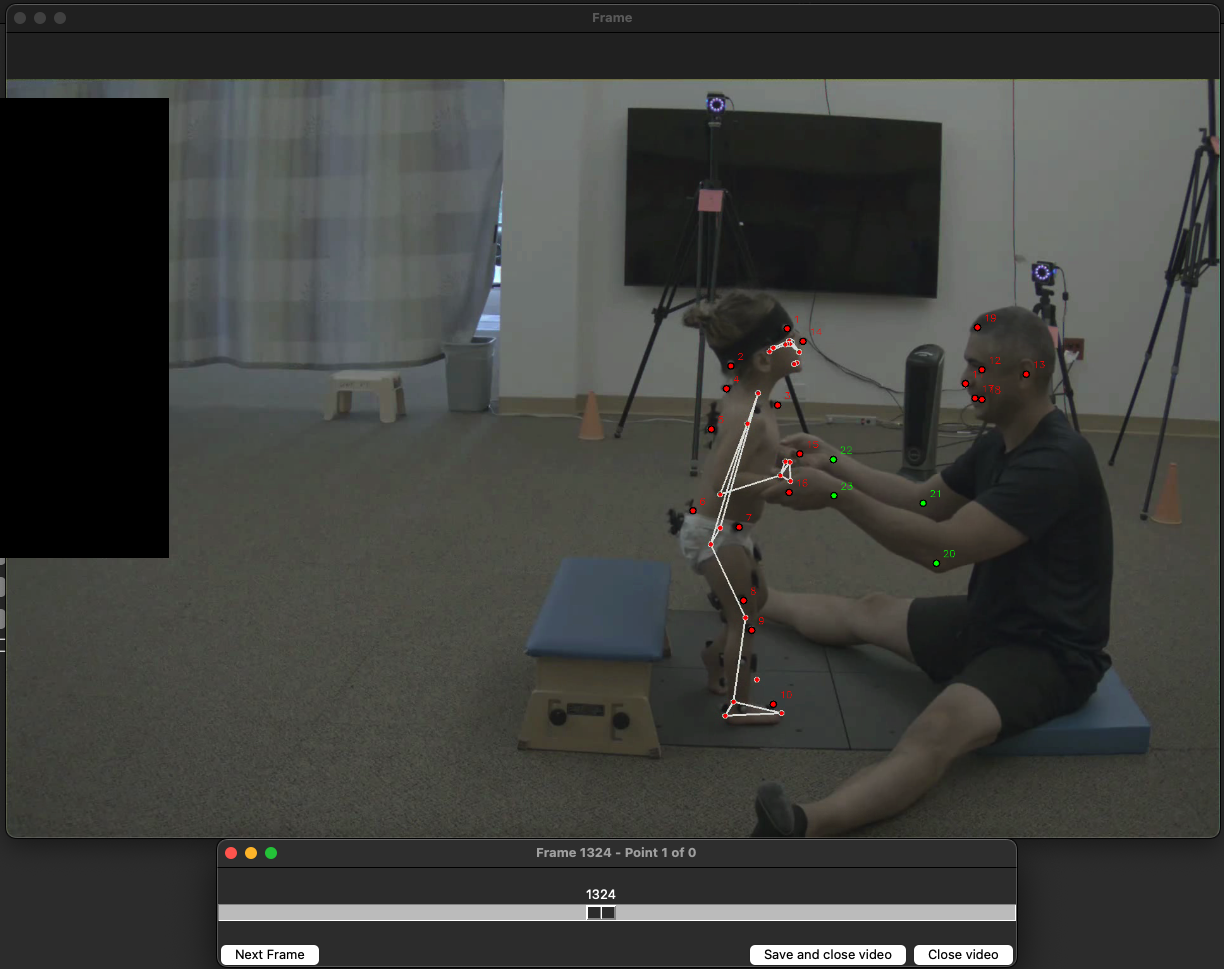}
\caption{Screenshot of the \textit{Get Pixel Coordinate} tool within \textit{vailá}. The interface allows users to navigate video frames, zoom in for precise point selection, and mark or adjust keypoints. This tool facilitates manual annotation and verification of keypoints in video data, enhancing the flexibility of the \textit{vailá} toolbox.}
\label{fig:getpixel}
\end{figure}

An example of the tool's interface is shown in Figure~\ref{fig:getpixel}. The interface is designed for efficient annotation, enabling users to interactively navigate through frames and adjust accurately to keypoints. This flexibility enhances the usability of the \textit{vailá} toolbox for analyzing biomechanical data from markerless systems.

The \textit{Get Pixel Coordinate} tool also integrates with the \textit{Edit CSV} tool within \textit{vailá}, which provides functionality for processing and converting keypoints or markers from various CSV formats. This integration enables users to seamlessly incorporate manually annotated data or adjust existing datasets for compatibility with other analysis workflows.

Incorporating tools like \texttt{getpixelvideo.py}, \textit{vailá} enhances the ability to process and analyze markerless video data, which is increasingly important in biomechanical studies involving participants who may not tolerate traditional markers or in field-based research where markerless systems are more practical. Using these workflows, the ability to quickly process large batches of video data allows researchers to adapt to unexpected research opportunities without losing valuable data collections.

Furthermore, the inclusion of such tools demonstrates the modularity and extensibility of \textit{vailá}. Users can customize and expand the toolbox by adding scripts tailored to their specific needs, promoting a collaborative environment where new functionalities can be shared and integrated into the main framework. This flexibility is crucial for accommodating the diverse requirements of biomechanics research, where data types and analysis methods can vary widely between studies.

Another significant advantage of \textit{vailá} is its cross-platform compatibility, particularly in providing tools for visualizing C3D files on macOS and Linux. Currently, most free or open-source tools for C3D visualization, such as Mokka and Visual3D Reader, are limited to Windows. While Mokka is available for macOS, it only runs on older OSX versions, leaving many users without accessible options.

To address this, \textit{vailá} includes powerful 3D visualization tools, allowing users to animate and interact with C3D and CSV files using Python libraries such as Matplotlib and Plotly. These visualizations can be displayed in the system's default browser, offering features like marker selection and forward/backward playback controls. This functionality enhances the user experience by providing interactive, real-time visualizations on any operating system.

The visualization tools are easily accessible under the "Available Tools > Visualization" section within the \textit{vailá} interface, enabling users to generate 3D plots and explore biomechanical data without the need for specialized, platform-specific software.

\begin{figure}[htbp]
\centering
\includegraphics[width=0.8\textwidth]{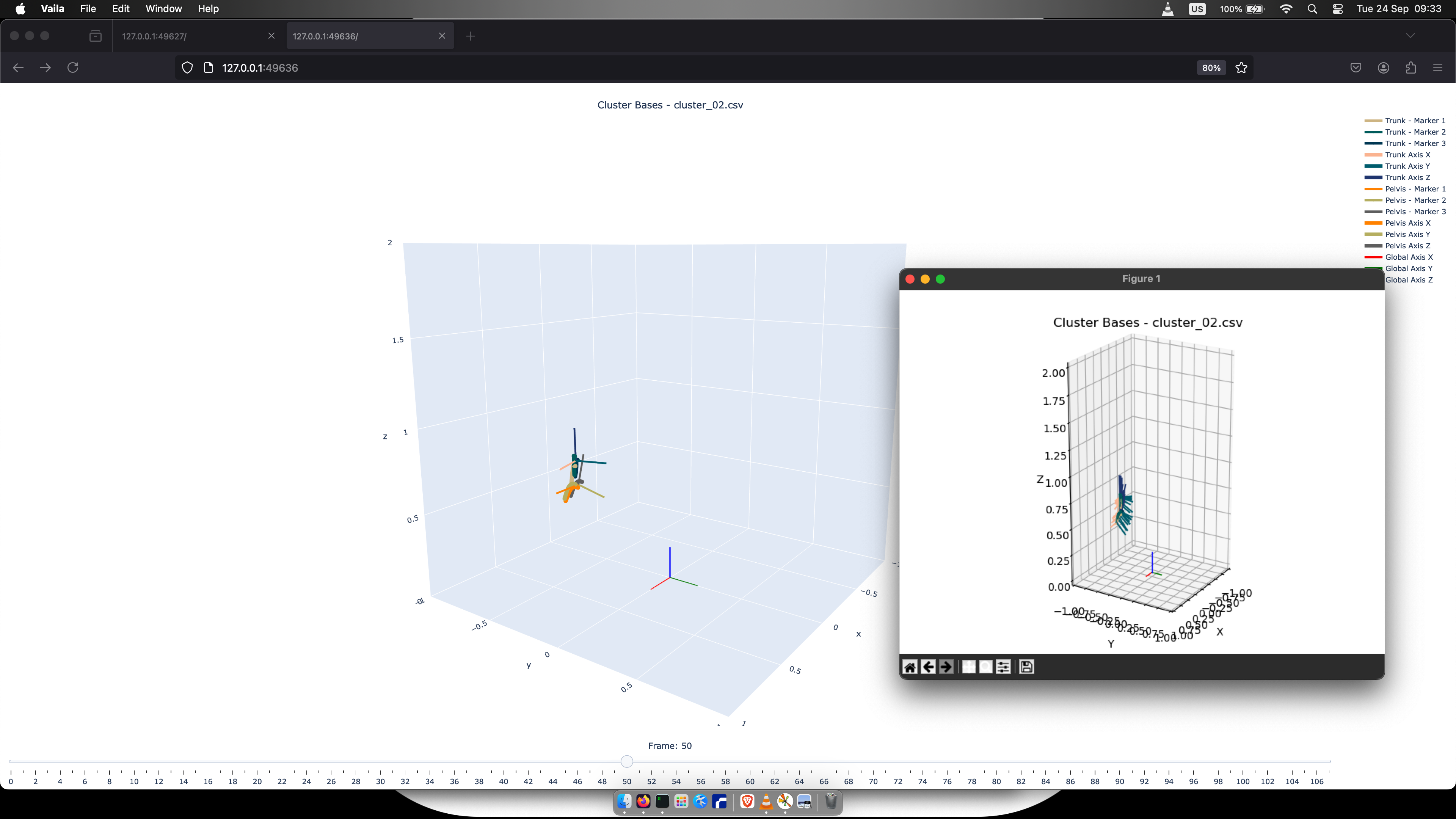}
\caption{Example of a 3D animated visualization of C3D data using Plotly in \textit{vailá}. The interface allows users to interact with markers, play the animation forward or backward, and adjust the display animation. These tools are located in the "Available Tools > Visualization" section.}
\label{fig:plotly}
\end{figure}

An example of a 3D animated C3D data visualization using Plotly within \textit{vailá} is shown in Figure~\ref{fig:plotly}.

\subsection{Markerless 2D Workflow in \textit{vailá}}

The \textit{vailá} toolbox provides an optimized and efficient workflow for processing markerless 2D video data, significantly improving performance compared to traditional methods. The workflow is represented in Figure~\ref{fig:markerless_2D_workflow_horizontal}, which shows the sequence of steps used in the processing pipeline:

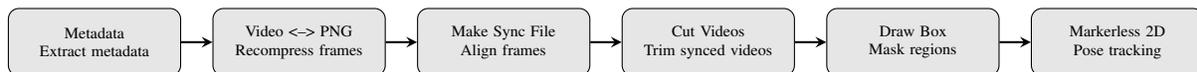
\begin{figure}[H]
    \centering
    \begin{tikzpicture}[node distance=3.2cm, every node/.style={scale=0.85}] 
        \tikzstyle{block} = [rectangle, draw, fill=gray!20, 
                             text width=7em, text centered, rounded corners, minimum height=3em, font=\scriptsize] 
        \tikzstyle{arrow} = [thick,->,>=stealth]
        
        \node (metadata) [block] {Metadata\\Extract metadata};
        \node (video2png) [block, right of=metadata] {Video <--> PNG\\Recompress frames};
        \node (sync) [block, right of=video2png] {Make Sync File\\Align frames};
        \node (cut) [block, right of=sync] {Cut Videos\\Trim synced videos};
        \node (drawbox) [block, right of=cut] {Draw Box\\Mask regions};
        \node (markerless) [block, right of=drawbox] {Markerless 2D\\Pose tracking};

        \draw [arrow] (metadata) -- (video2png);
        \draw [arrow] (video2png) -- (sync);
        \draw [arrow] (sync) -- (cut);
        \draw [arrow] (cut) -- (drawbox);
        \draw [arrow] (drawbox) -- (markerless);
    \end{tikzpicture}
    \caption{Markerless 2D processing pipeline in \textit{vailá}, with each step briefly describing the operation performed.}
    \label{fig:markerless_2D_workflow_horizontal}
\end{figure}

The workflow starts by extracting metadata from the video using the \textit{Metadata info} tool. This step identifies any potential video compression or codec issues before further processing.

The \textit{Video <--> PNG} tool is applied to extract all frames from the original video and recompress them into a new h264 video format, which is optimal for analysis. Next, the \textit{Make Sync File} tool generates synchronization files to align the frames of two videos for subsequent analysis.

The \textit{Cut Videos} step trims the synchronized videos using the sync file, and the \textit{Draw Box} tool masks irrelevant regions in the video, focusing on the subject of interest. Finally, the \textit{Markerless 2D} Multimodal, powered by MediaPipe~\citep{lugaresi2019mediapipe}, tracks and overlays 2D pose landmarks on the processed video.

This workflow generates the following output files for each processed video:

\begin{itemize}
    \item \textbf{Processed Video} (\texttt{*\_mp.mp4}): The video with 2D pose landmarks overlaid on the original frames.
    \item \textbf{Normalized Landmark CSV} (\texttt{*\_mp\_norm.csv}): A CSV file containing normalized landmark coordinates.
    \item \textbf{Pixel Landmark CSV} (\texttt{*\_mp\_pixel.csv}): A CSV file containing pixel coordinates for the landmarks, scaled to the video’s resolution.
    \item \textbf{Log File} (\texttt{log\_info.txt}): A log file detailing video metadata, such as resolution, frame rate, total frames, and codec used.
\end{itemize}

\begin{figure}[htbp]
\centering
    \begin{minipage}[t]{0.48\textwidth}
        \centering
        \includegraphics[width=0.98\textwidth]{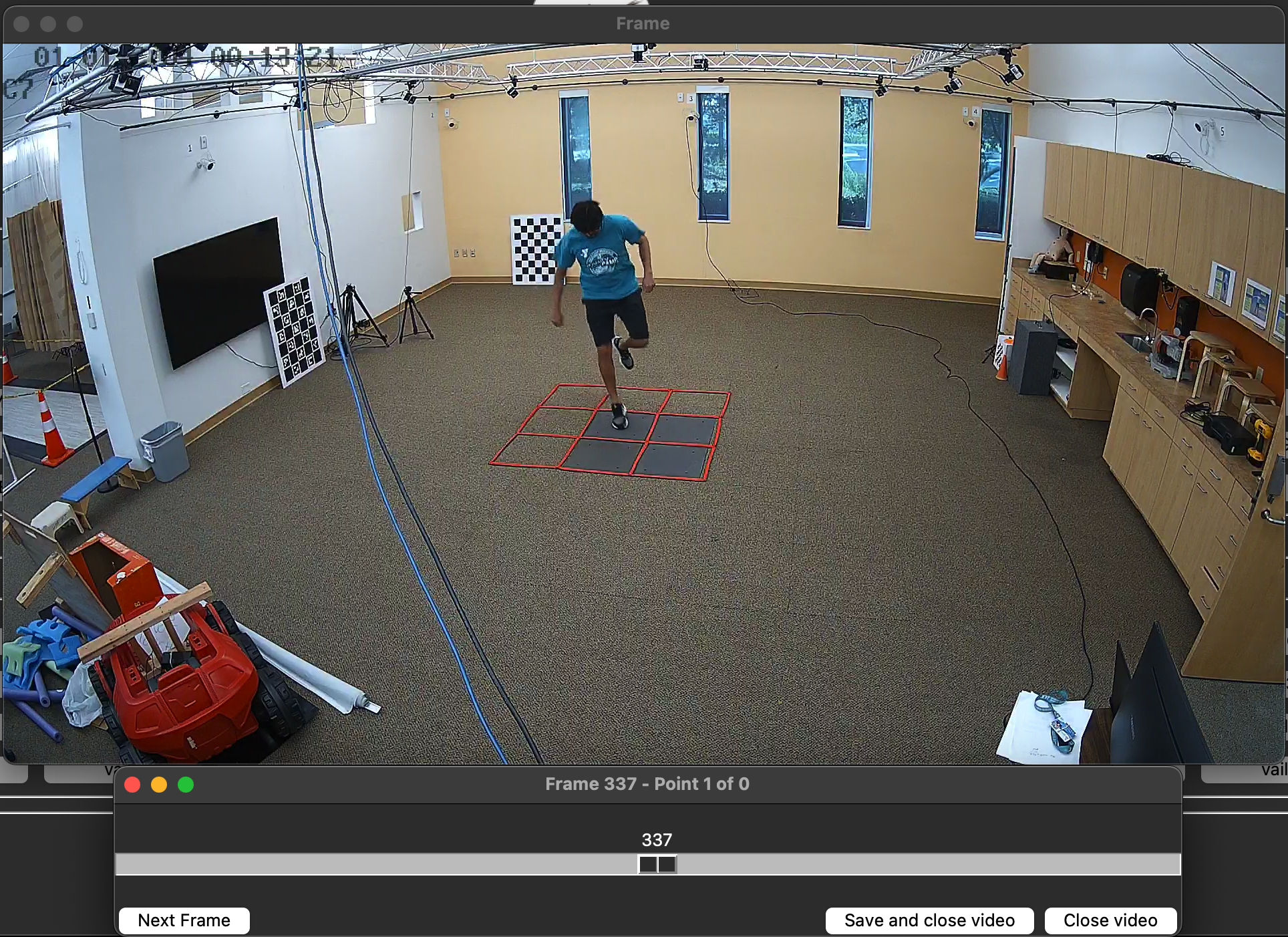}
        \subcaption{(a) Raw video camera 1}\label{fig1_cube_markerless}
    \end{minipage}
    \hfill
    \begin{minipage}[t]{0.48\textwidth}
        \centering
        \includegraphics[width=0.98\textwidth]{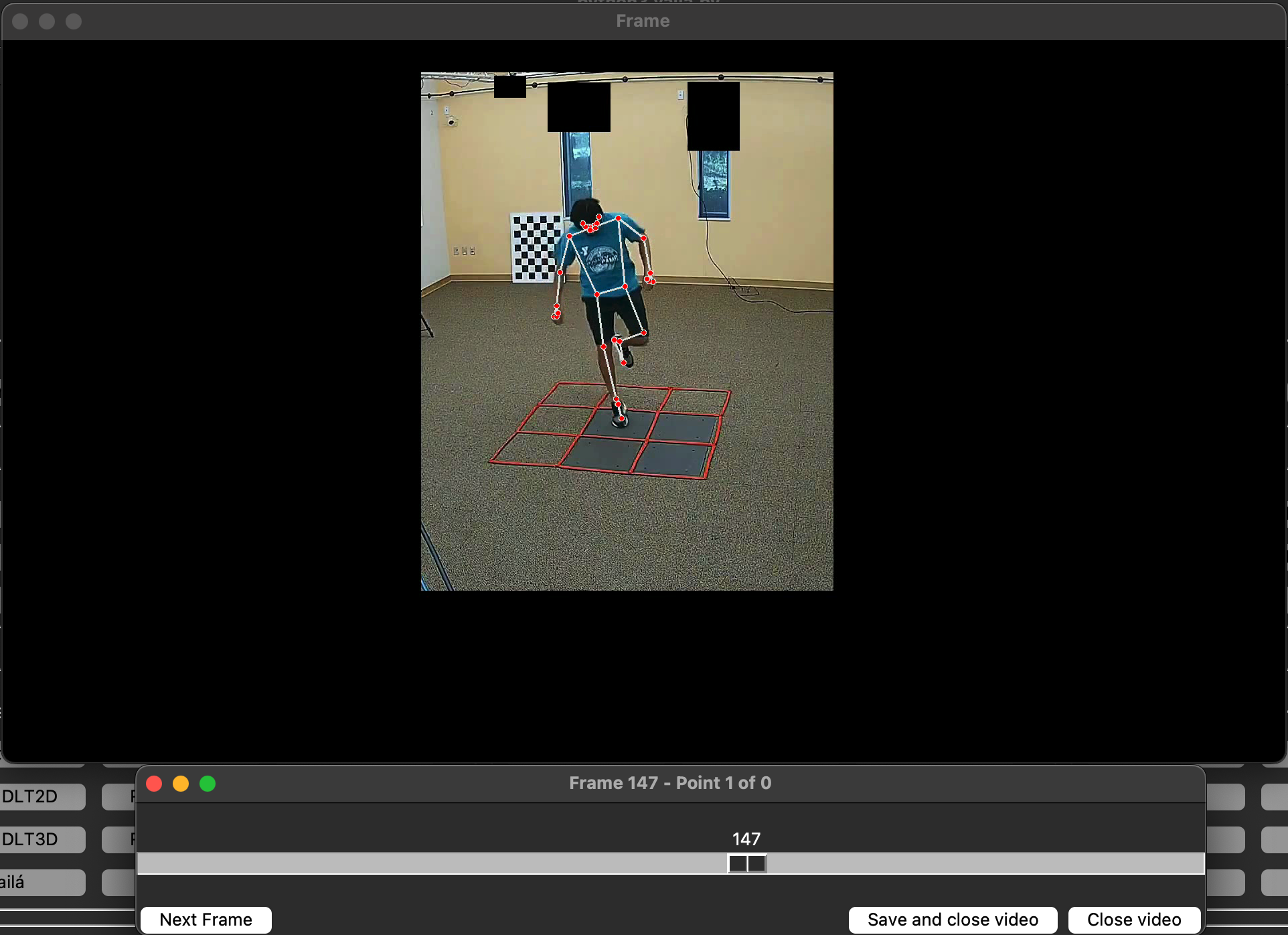}
        \subcaption{(b) \textit{vailá} processed video of camera 1 with Landmarks}\label{fig2_cube_markerless}
    \end{minipage}
    
    \vspace{0.5cm} 
    
    \begin{minipage}[t]{0.48\textwidth}
        \centering
        \includegraphics[width=0.98\textwidth]{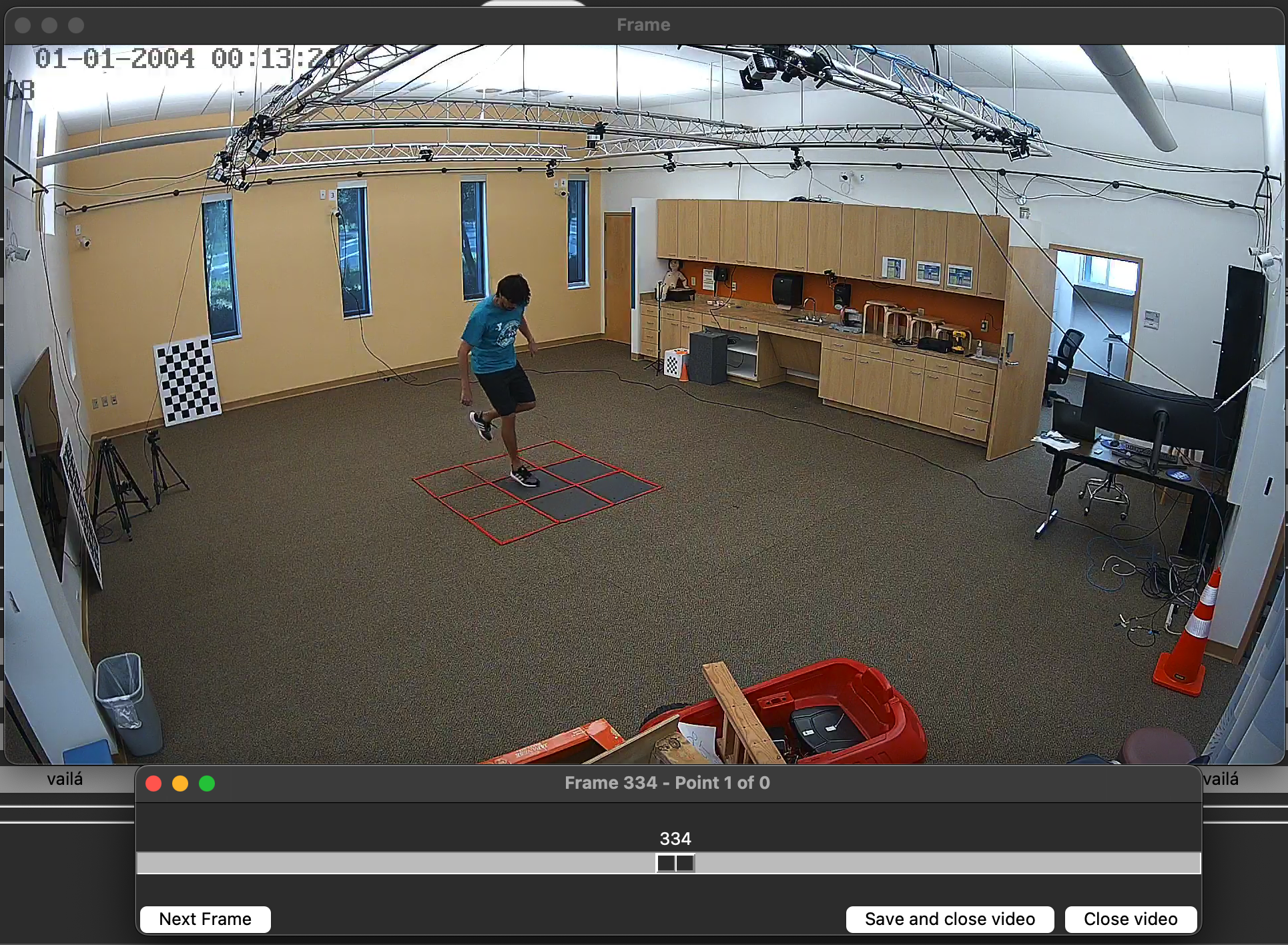}
        \subcaption{(c) Raw video camera 2}\label{fig3_cube_markerless}
    \end{minipage}
    \hfill
    \begin{minipage}[t]{0.48\textwidth}
        \centering
        \includegraphics[width=0.98\textwidth]{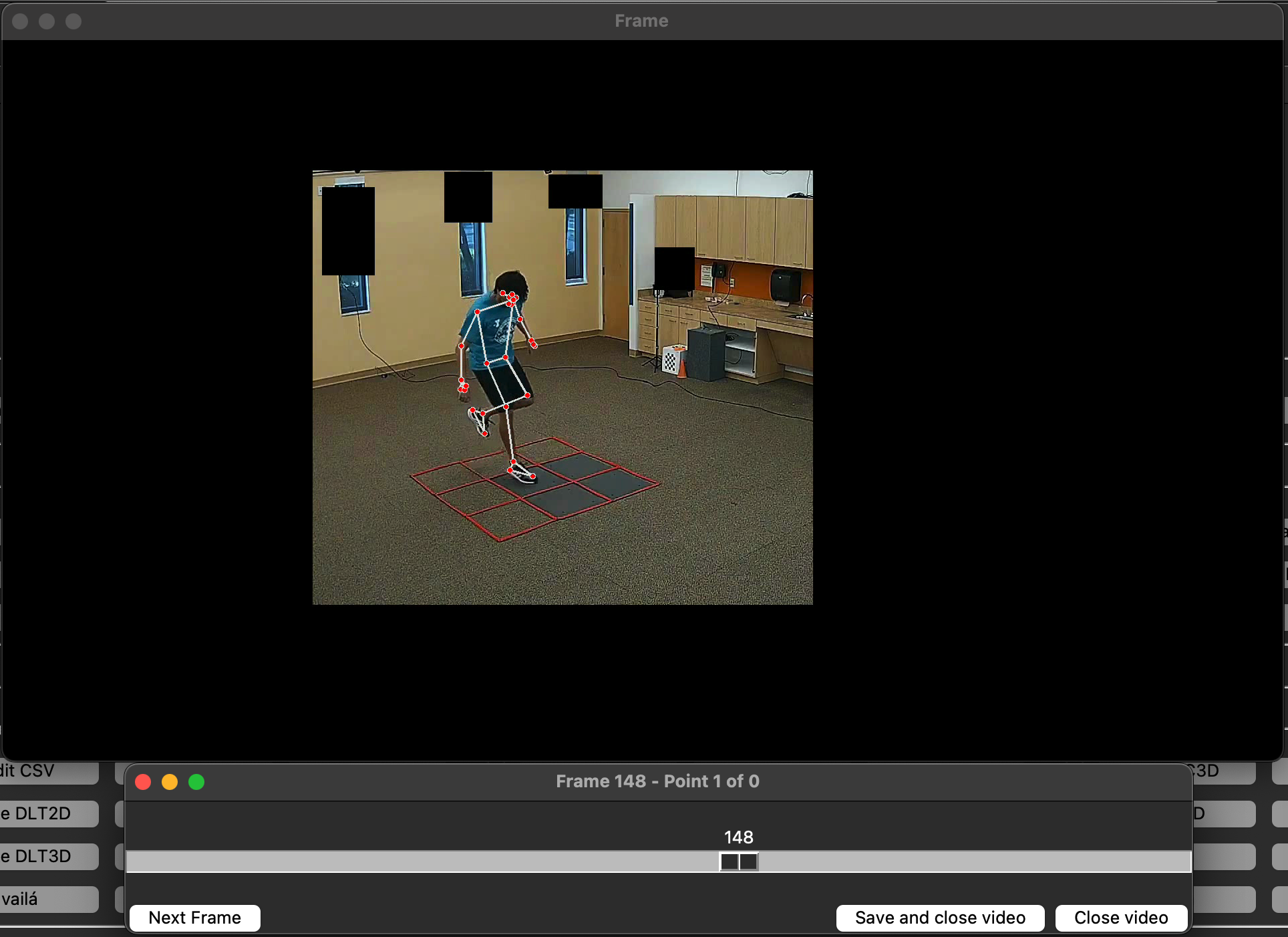}
        \subcaption{(d) \textit{vailá} processed video of camera 2 with Landmarks}\label{fig4_cube_markerless}
    \end{minipage}
    
    \vspace{0.5cm} 
    
    \caption{Markerless 2D analysis in \textit{vailá}. Top: raw video frames. Bottom: processed videos with 2D pose landmarks overlaid.}
    \label{fig:markerless_mosaic}
\end{figure}

The Figure~\ref{fig:markerless_mosaic} presents a mosaic of four frames from the markerless workflow to visually demonstrate the results. The top-left (Figure~\ref{fig:markerless_mosaic}a) shows the raw input video from Camera 1, and the top-right (Figure~\ref{fig:markerless_mosaic}b) displays the processed video with the 2D pose landmarks overlaid. Similarly, the bottom-left (Figure~\ref{fig:markerless_mosaic}c) shows the raw video from Camera 2, and the bottom-right (Figure~\ref{fig:markerless_mosaic}d) displays the final processed result with landmarks.

In addition to the markerless workflow, \textit{vailá} also includes tools for creating 2D and 3D calibrations using the Direct Linear Transformation (DLT) method, both for static and dynamic (moving) cameras. These scripts are available under \textit{Available Tools} > \textit{Data Files} > \textit{Make DLT 2D} and \textit{Make DLT 3D}. Moreover, 2D and 3D reconstructions based on single or multiple DLT calibrations for each frame are supported, allowing dynamic calibrations. These tools can be used with screen coordinate data, either from the Markerless 2D system or any other tracking system.

\subsection{Motion Capture and IMU Cluster Workflows in \textit{vailá}}

Using a cluster-based approach, the \textit{vailá} toolbox offers robust workflows for analyzing motion capture and IMU (Inertial Measurement Unit) data. These techniques allow the processing of 3D coordinates and sensor data to compute joint rotations and derive Euler angles. The two modalities—traditional motion capture clusters and IMU-based clusters—share a similar analysis pipeline but differ in the data sources used.

\subsubsection{Motion Capture Cluster Workflow}

In the motion capture cluster workflow, 3D coordinates of reflective markers attached to specific clusters (e.g., trunk and pelvis) are processed to compute orthonormal bases and derive Euler angles that describe joint rotations. The workflow organizes the outputs into a structured directory, including folders for figures and processed data files.

The output includes:
\begin{itemize}
    \item \textbf{Cluster Data CSV} (\texttt{*\_cluster.csv}): This file contains the computed Euler angles for each cluster over time.
    \item \textbf{Figures} (\texttt{*\_figure.png}): Visual representations of the orthonormal bases and Euler angles for each cluster.
\end{itemize}

\begin{figure}[htbp]
    \centering
    \begin{minipage}[t]{0.48\textwidth}
        \centering
        \includegraphics[width=0.95\textwidth]{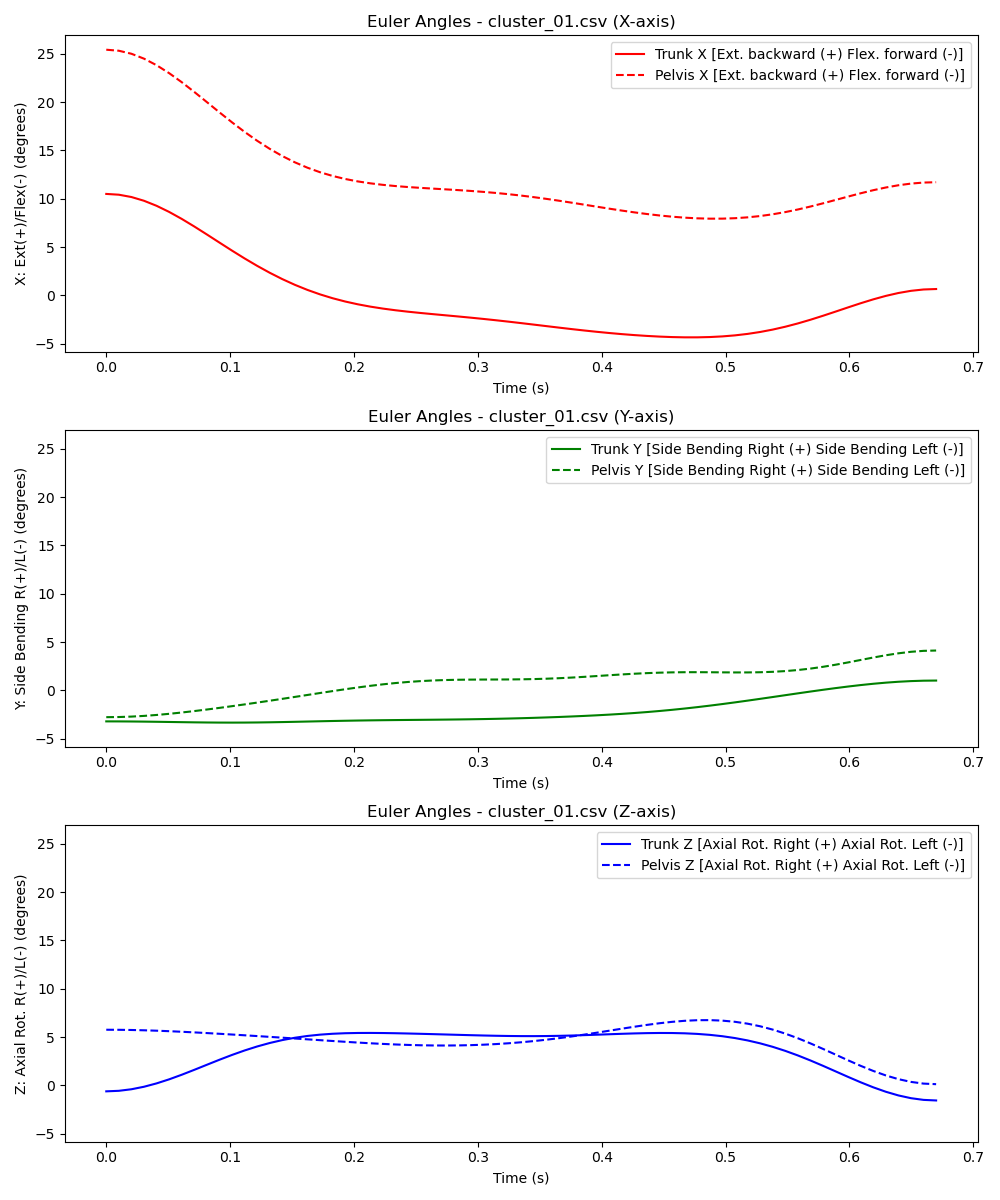}
        \subcaption{Euler Angles for Trunk (Cluster 1) - Typically Developing Child}\label{fig:euler_cluster_a}
    \end{minipage}
    \hfill
    \begin{minipage}[t]{0.48\textwidth}
        \centering
        \includegraphics[width=0.95\textwidth]{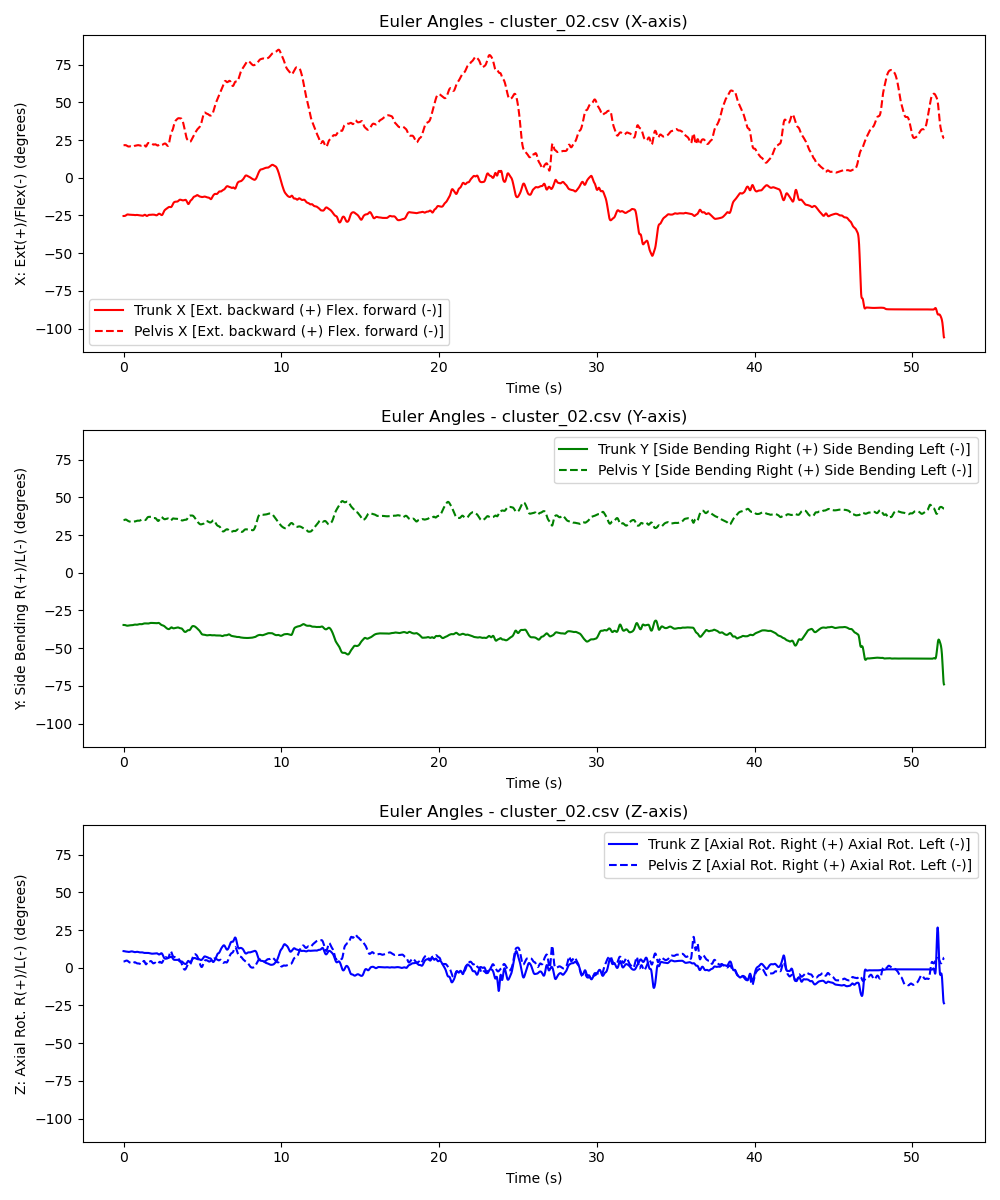}
        \subcaption{Euler Angles for Pelvis (Cluster 2) - Child with Cerebral Palsy}\label{fig:euler_cluster_b}
    \end{minipage}
    \caption{Euler angles over time for the trunk and pelvis clusters during a Sit To Stand Test in \textit{vailá}. Clusters were positioned on the trunk and pelvis. (a) Data from a typically developing child. (b) Data from a child with cerebral palsy. The use of clusters is beneficial when skin issues or altered sensitivity may lead to children resisting the placement of kinematic markers.}
    \label{fig:euler_angles_results}
\end{figure}

The results of this workflow are demonstrated in Figure~\ref{fig:euler_angles_results}, showing the Euler angles for both the trunk (Cluster 1) and pelvis (Cluster 2) during a Sit To Stand Test.

\subsubsection{IMU Cluster Workflow}

The IMU cluster workflow in \textit{vailá} offers a similar analysis structure but utilizes data from Inertial Measurement Units (IMUs) instead of reflective markers. IMUs are typically placed on the same body regions as the clusters (e.g., trunk and pelvis) and provide accelerometer and gyroscope data along the X, Y, and Z axes. Euler angles are derived from this data to describe the same joint rotations computed in the motion capture workflow.

The output of the IMU workflow is stored in a CSV file with the following structure:
\begin{scriptsize}
\begin{verbatim}
Time,Gyro_X,Gyro_Y,Gyro_Z,Acc_X,Acc_Y,Acc_Z,Euler_X,Euler_Y,Euler_Z, Tilt_X,Tilt_Y,Tilt_Z,Quat_W,Quat_X,Quat_Y,Quat_Z
0.000,52.874,17.166,0.961,-0.066,0.910,0.858,0.235,0.024,0.000,-3.038,46.609,46.770,1.000,0.002,0.000,0.000
...
10.000,52.872,17.166,0.962,-0.066,0.910,0.858,0.468,0.047,0.000,-3.038,46.608,46.770,0.999,0.004,0.000,-0.000
\end{verbatim}
\end{scriptsize}

This file contains timestamps and a range of computed values, including:
\begin{itemize}
    \item \textbf{Gyroscope Data} (X, Y, Z): Captures the angular velocity of the IMU in three axes.
    \item \textbf{Accelerometer Data} (X, Y, Z): Measures the linear acceleration of each IMU in three axes.
    \item \textbf{Euler Angles} (X, Y, Z): Derived angles representing the rotation of the body segments.
    \item \textbf{Tilt Angles} (X, Y, Z): Calculations of the tilt of each segment.
    \item \textbf{Quaternion Data} (W, X, Y, Z): Provides an alternative representation of the orientation of each segment.
\end{itemize}

The key difference between the motion capture cluster and IMU workflows lies in the data sources: while motion capture uses physical markers and camera systems, the IMU workflow leverages wearable sensors to gather acceleration and angular velocity data. This allows for more flexible motion tracking, particularly in environments where camera systems may not be feasible or when working with subjects who may resist wearing reflective markers.

The following figure compares Euler angles derived from a typically developing child and a child with cerebral palsy, showcasing how both modalities can assess joint movements across different populations and conditions.

\subsection{EMG Workflow in \textit{vailá}}

\begin{figure}[htbp]
    \centering
    \begin{tikzpicture}[node distance=4.5cm, every node/.style={scale=0.85}]
        \tikzstyle{block} = [rectangle, draw, fill=gray!20, 
                             text width=8em, text centered, rounded corners, minimum height=3em]
        \tikzstyle{arrow} = [thick,->,>=stealth]
        
        \node (selectpath) [block] {Select path of files CSV\\Choose directory with EMG data};
        \node (defineparams) [block, right of=selectpath] {Define parameters\\Sampling rate, etc.};
        \node (checkwindows) [block, right of=defineparams] {Check analysis window\\Specify start and end};
        \node (done) [block, right of=checkwindows] {Results and figures generated};

        \draw [arrow] (selectpath) -- (defineparams);
        \draw [arrow] (defineparams) -- (checkwindows);
        \draw [arrow] (checkwindows) -- (done);
    \end{tikzpicture}
    \caption{EMG signal processing workflow in \textit{vailá}.}
    \label{fig:emg_workflow}
\end{figure}
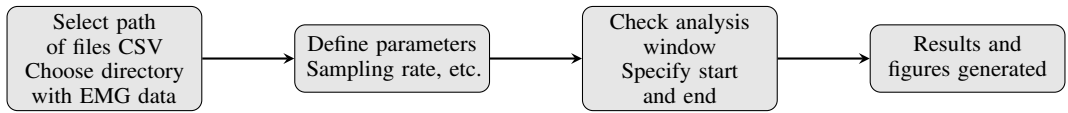

The processing workflow for EMG data in \textit{vailá} begins with selecting the directory containing CSV files with EMG data, defining the sampling rate, and identifying the analysis windows. The key outputs include filtered EMG signals, rectified signals, RMS values, and frequency domain analysis, which are visualized in Figure~\ref{fig:emg_workflow}.

\begin{figure}[htbp]
    \centering
    \begin{minipage}[t]{0.95\textwidth} 
        \centering
        \includegraphics[width=0.95\textwidth]{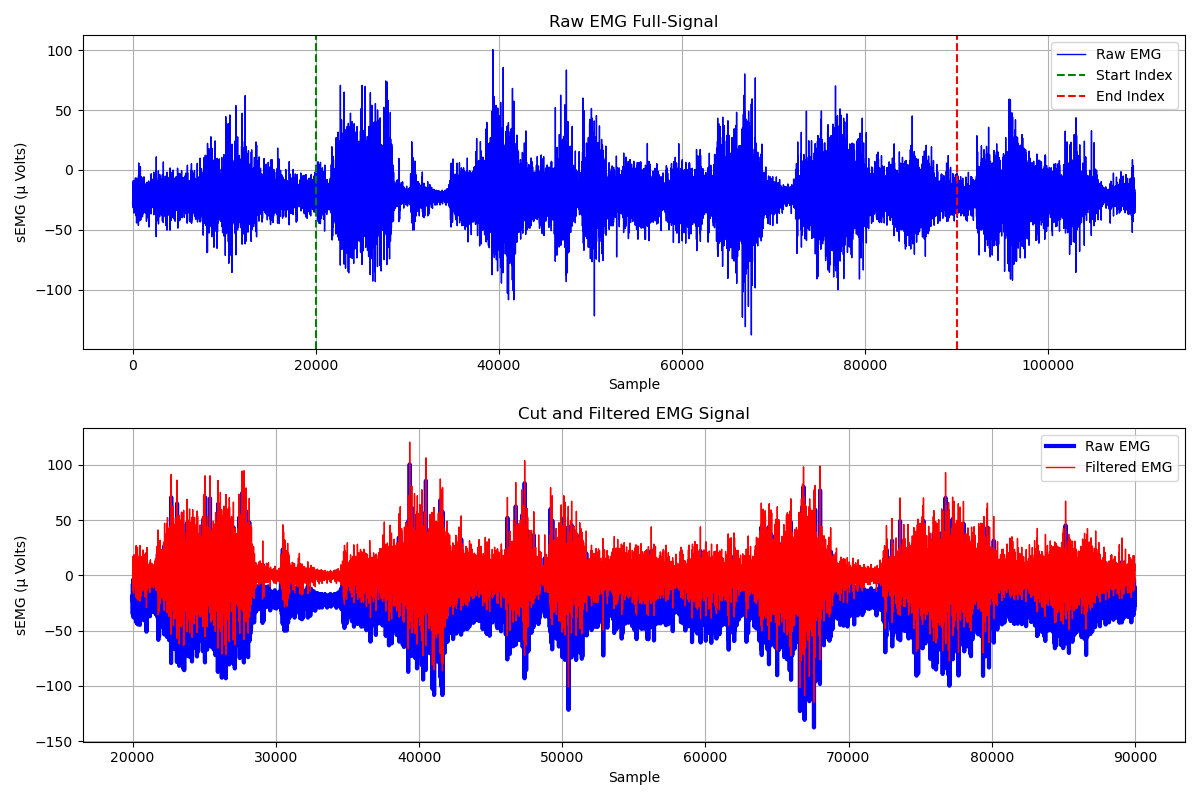}
        \subcaption{Filtered EMG signal}\label{fig:emg_filtered}
    \end{minipage}
    \vspace{0.3cm} 

    \begin{minipage}[t]{0.45\textwidth} 
        \centering
        \includegraphics[width=0.95\textwidth]{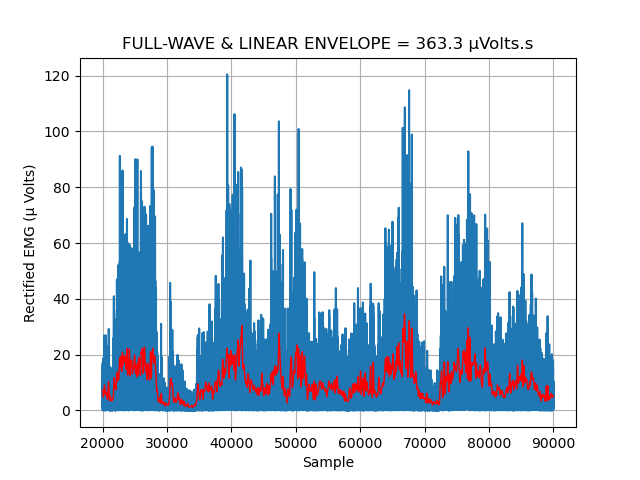}
        \subcaption{Rectified EMG}\label{fig:emg_rectified}
    \end{minipage}
    \hfill
    \begin{minipage}[t]{0.45\textwidth}
        \centering
        \includegraphics[width=0.95\textwidth]{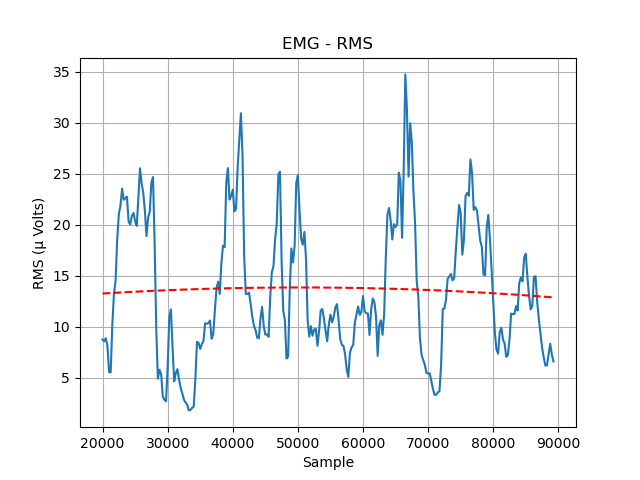}
        \subcaption{RMS values}\label{fig:emg_rms}
    \end{minipage}
    
    \vspace{0.3cm} 
    \begin{minipage}[t]{0.45\textwidth}
        \centering
        \includegraphics[width=0.95\textwidth]{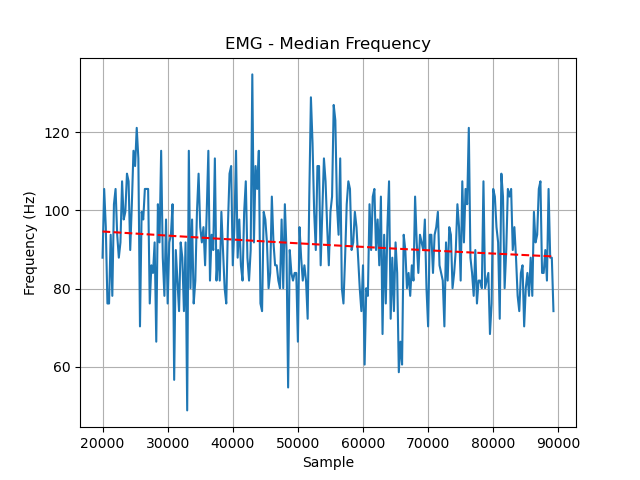}
        \subcaption{Median Frequency}\label{fig:emg_median_frequency}
    \end{minipage}
    \hfill
    \begin{minipage}[t]{0.45\textwidth}
        \centering
        \includegraphics[width=0.95\textwidth]{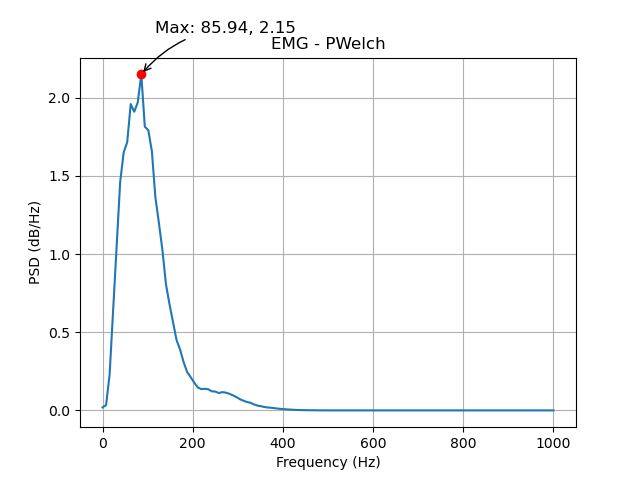}
        \subcaption{Welch's PSD}\label{fig:emg_pwelch}
    \end{minipage}

    \caption{Key results of EMG analysis from \textit{vailá} for file \texttt{emg\_01.csv}. Top: filtered EMG signal. Bottom: rectified signal, RMS values, median frequency, and Welch's PSD.}
    \label{fig:emg_mosaic}
\end{figure}

To demonstrate the results of EMG analysis, Figure~\ref{fig:emg_mosaic} presents a mosaic of key outputs for the file \texttt{emg\_01.csv}. These include the bandpass filtered EMG signal, which is used to remove noise; the full-wave rectified EMG signal for envelope detection; the Root Mean Square (RMS) values, which reflect muscle activity; the median frequency, showing trends over time and indicating muscle fatigue; and Welch's Power Spectral Density (PSD), which highlights the frequency components of the signal.

\subsection{Force Cube Analysis in \textit{vailá}}

The \textit{vailá} system supports a robust workflow for analyzing force platform data using the Force Cube method. This process involves analyzing the vertical ground reaction force (Fz) and calculating key biomechanical metrics, such as peak forces, rate of force development (RFD), and stiffness. The system is designed to allow batch processing of multiple files, and each analysis is saved automatically.

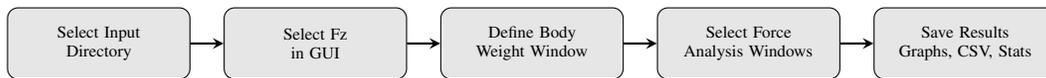
\begin{figure}[H]
    \centering
    \begin{tikzpicture}[node distance=3.2cm, every node/.style={scale=0.9}]
        \tikzstyle{block} = [rectangle, draw, fill=gray!20, 
                             text width=7em, text centered, rounded corners, minimum height=3em, font=\scriptsize]
        \tikzstyle{arrow} = [thick,->,>=stealth]
        
        \node (selectinput) [block] {Select Input\\Directory};
        \node (selectfz) [block, right of=selectinput] {Select Fz\\in GUI};
        \node (definebw) [block, right of=selectfz] {Define Body\\Weight Window};
        \node (selectwindows) [block, right of=definebw] {Select Force\\Analysis Windows};
        \node (save) [block, right of=selectwindows] {Save Results\\Graphs, CSV, Stats};

        \draw [arrow] (selectinput) -- (selectfz);
        \draw [arrow] (selectfz) -- (definebw);
        \draw [arrow] (definebw) -- (selectwindows);
        \draw [arrow] (selectwindows) -- (save);
    \end{tikzpicture}
    \caption{Force Cube Analysis workflow in \textit{vailá}, outlining each step from input directory selection to saving results.}
    \label{fig:force_cube_workflow}
\end{figure}

The Force Cube Analysis workflow consists of several steps described in Figure~\ref{fig:force_cube_workflow}.

The following Figure~\ref{fig:force_cube_combined} illustrates the peak selection process (a) and the generated force-time curve (b), which are key outputs of the Force Cube Analysis.

\begin{figure}[htbp]
    \centering
    \begin{minipage}[t]{0.45\textwidth}
        \centering
        \includegraphics[width=0.95\textwidth]{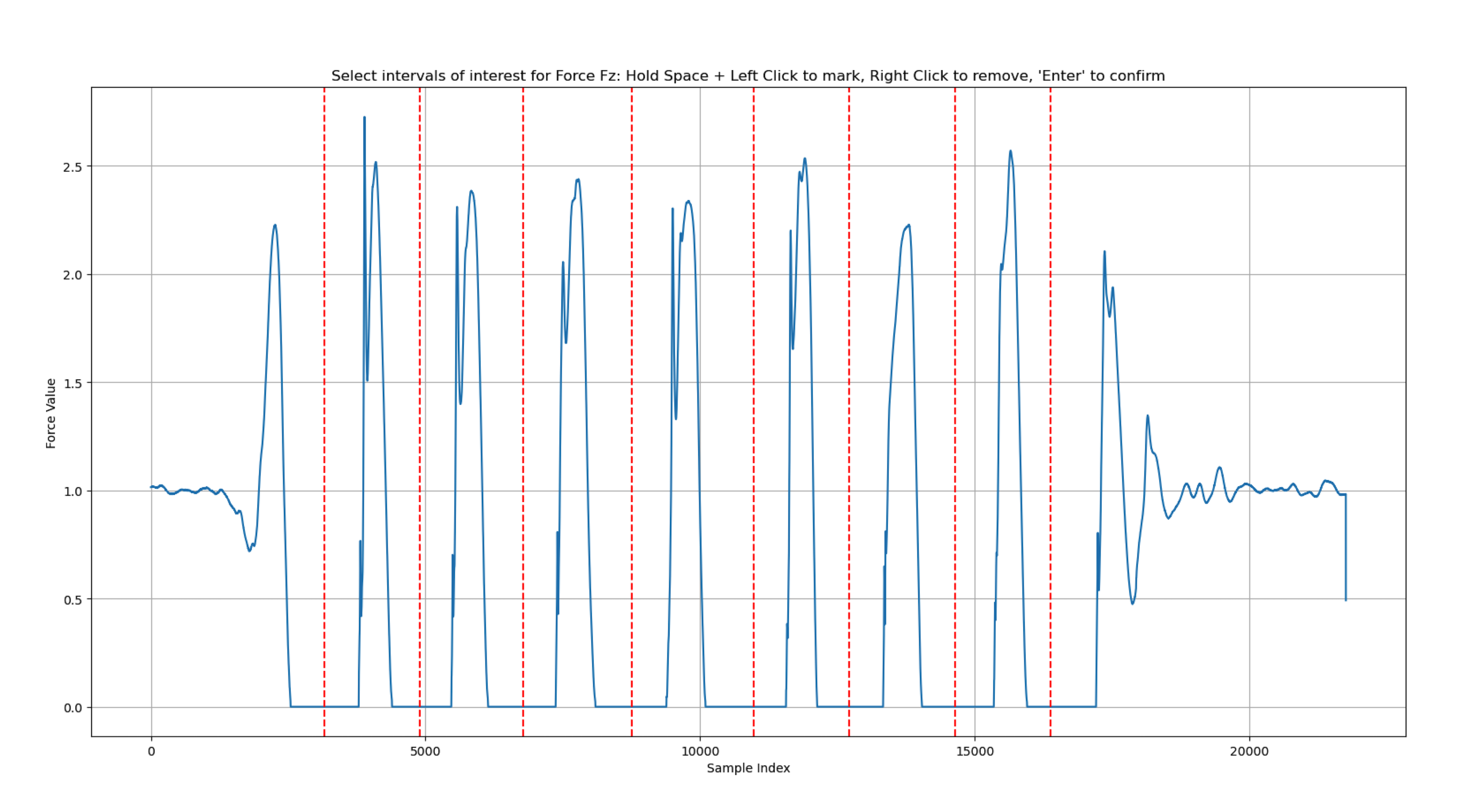}
        \subcaption{Peak selection for analysis.}\label{fig:select_peaks}
    \end{minipage}%
    \hfill
    \begin{minipage}[t]{0.45\textwidth}
        \centering
        \includegraphics[width=0.95\textwidth]{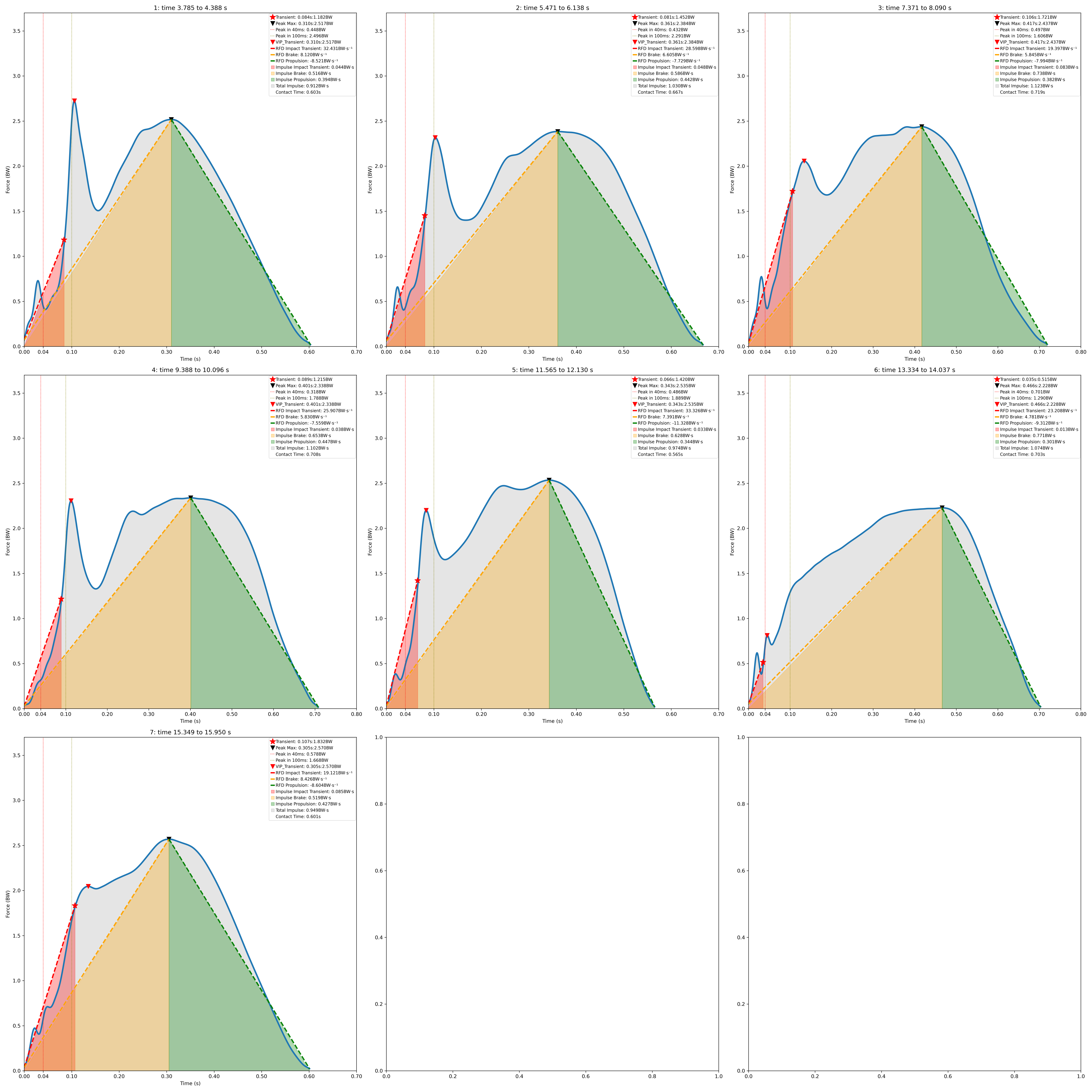}
        \subcaption{Force-time curve and biomechanical parameters.}\label{fig:force_cube_output}
    \end{minipage}
    
    \caption{Force Cube Analysis outputs: (a) peak selection and (b) force-time curve.}
    \label{fig:force_cube_combined}
\end{figure}

Each output file generated by the Force Cube Analysis contains key biomechanical metrics that are crucial for understanding the force dynamics during specific time windows. The metrics include: \\ 
File name (\texttt{FileName}), the date and time of the analysis (\texttt{TimeStamp}), the trial number (\texttt{Trial}), and the body weight in kilograms (\texttt{BW\_kg}). The analysis also records which foot was used (right or left) (\texttt{SideFoot\_RL}) and foot dominance (\texttt{Dominance\_RL}). A user-provided quality rating (\texttt{Quality}) and the number of samples in the trial (\texttt{Num\_Samples}) are also included. Indices for specific time points such as 40 ms (\texttt{Index\_40ms}), 100 ms (\texttt{Index\_100ms}), the impact transient (\texttt{Index\_ITransient}), the vertical impact peak (\texttt{Index\_VIP}), and the maximum force peak (\texttt{Index\_Max}) are provided. The trial’s duration (\texttt{Test\_Duration\_s}), cumulative time sum (\texttt{CumSum\_Times\_s}), contact time (\texttt{Contact\_Time\_s}), and specific times for 40 ms, 100 ms, impact transient, vertical impact peak, and maximum force (\texttt{Time\_40ms\_s}, \texttt{Time\_100ms\_s}, \texttt{Time\_ITransient\_s}, \texttt{Time\_VIP\_s}, and \texttt{Time\_Peak\_VMax\_s}) are also included. Key force-related metrics such as peak forces at different times (normalized to body weight) are recorded: \texttt{VPeak\_40ms\_BW}, \texttt{VPeak\_100ms\_BW}, \texttt{Peak\_VITransient\_BW}, \texttt{Peak\_VIP\_BW}, and \texttt{Peak\_VMax\_BW}. The total impulse (\texttt{Total\_Imp\_BW.s}) and impulse at various stages (\texttt{Imp\_40ms\_BW.s}, \texttt{Imp\_100ms\_BW.s}, \texttt{Imp\_ITransient\_BW.s}, \texttt{Imp\_Brake\_VMax\_BW.s}, \texttt{Imp\_Propulsion\_BW.s}) are calculated, alongside the rate of force development (RFD) at different intervals (\texttt{RFD\_40ms\_BW.s\textsuperscript{-1}}, \texttt{RFD\_100ms\_BW.s\textsuperscript{-1}}, \texttt{RFD\_ITransient\_BW.s\textsuperscript{-1}}, \texttt{RFD\_Brake\_VMax\_BW.s\textsuperscript{-1}}, and \texttt{RFD\_Propulsion\_BW.s\textsuperscript{-1}}). Finally, stiffness parameters (\texttt{Simple\_stiffness\_constant}, \texttt{High\_stiffness}, \texttt{Low\_stiffness}, \texttt{Transition\_time}) and the average loading rate (\texttt{Average\_loading\_rate}) complete the set of metrics.

\subsection{CoP Balance Analysis in \textit{vailá}}

\begin{figure}[H]
    \centering
    \begin{tikzpicture}[node distance=3.0cm, every node/.style={scale=0.8}]
        \tikzstyle{block} = [rectangle, draw, fill=gray!20, 
                             text width=7em, text centered, rounded corners, minimum height=3em, font=\scriptsize]
        \tikzstyle{arrow} = [thick,->,>=stealth]
        
        \node (forceplate) [block] {Force Plate};
        \node (copanalysis) [block, right of=forceplate] {CoP Balance\\Analysis};
        \node (selectinput) [block, right of=copanalysis] {Select Input\\Directory};
        \node (selectoutput) [block, right of=selectinput] {Select Output\\Directory};
        \node (selectcxcy) [block, right of=selectoutput] {Select Cx and Cy\\in GUI};
        \node (nextdefaults) [block, right of=selectcxcy] {Next in Default\\Parameters};
        \node (results) [block, right of=nextdefaults] {Results Data\\and Figures};

        \draw [arrow] (forceplate) -- (copanalysis);
        \draw [arrow] (copanalysis) -- (selectinput);
        \draw [arrow] (selectinput) -- (selectoutput);
        \draw [arrow] (selectoutput) -- (selectcxcy);
        \draw [arrow] (selectcxcy) -- (nextdefaults);
        \draw [arrow] (nextdefaults) -- (results);
    \end{tikzpicture}
    \caption{Workflow for CoP Balance Analysis in \textit{vailá}, excluding the confirmation step.}
    \label{fig:cop_workflow_horizontal_no_confirm}
\end{figure}
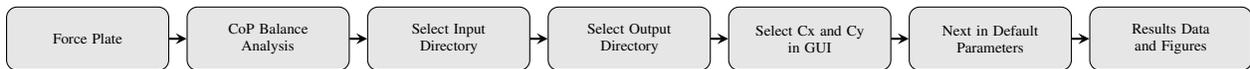

The \textit{vailá} system includes a comprehensive workflow for performing center of pressure (CoP) balance analysis based on force plate data. The analysis begins by selecting the CSV file directory with force plate data. After confirming the default parameters, the user selects the relevant CoP data columns, such as Cx (mediolateral) and Cy (anteroposterior) coordinates, through the graphical interface. Once confirmed, all figures and metrics are generated and saved automatically Figure~\ref{fig:cop_workflow_horizontal_no_confirm}.

The key outputs generated by \textit{vailá} include time-domain and frequency-domain metrics, as discussed in the literature by Duarte and Freitas \citep{duarte2010revision}, such as the CoP path length, mean speed, and power spectral density (PSD) for both mediolateral (ML) and anteroposterior (AP) directions. These metrics provide insights into the subject's balance stability and postural control.

The outputs generated from the CoP analysis include the following: an Overview Image, which provides a summary of key metrics such as total path length and mean speed; the Final Figure, showcasing detailed results including the CoP path plotted over time; a Metrics CSV File, which contains numerical values for key balance metrics such as mean ML and AP positions, RMS, and total path length; a PSD Plot, illustrating the power spectral density (PSD) and the frequency distribution of CoP movement; and a Stabilogram, depicting the CoP trajectory over time in both ML and AP directions.

The key balance metrics and frequency-domain characteristics (e.g., total power and frequency dispersion) are detailed in the CSV file. The results enable a robust posturographic analysis in line with the methodology proposed by Duarte and Freitas \citep{duarte2010revision}.

\begin{figure}[htbp]
    \centering
    \begin{minipage}[t]{0.45\textwidth}
        \centering
        \includegraphics[width=0.95\textwidth]{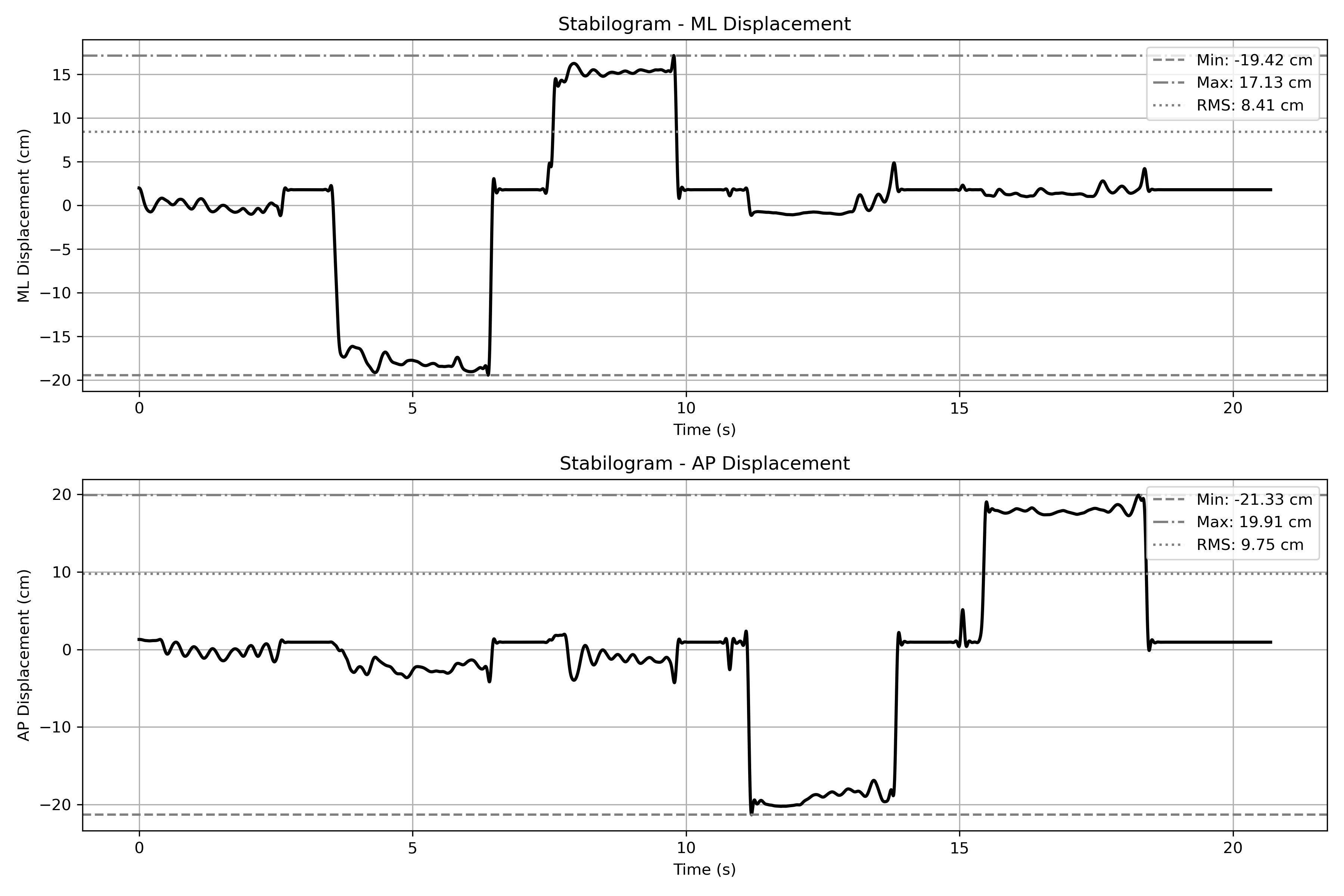}
        \subcaption{CoP stabilogram.}\label{fig:cop_stabilogram}
    \end{minipage}
    \hfill
    \begin{minipage}[t]{0.45\textwidth}
        \centering
        \includegraphics[width=0.95\textwidth]{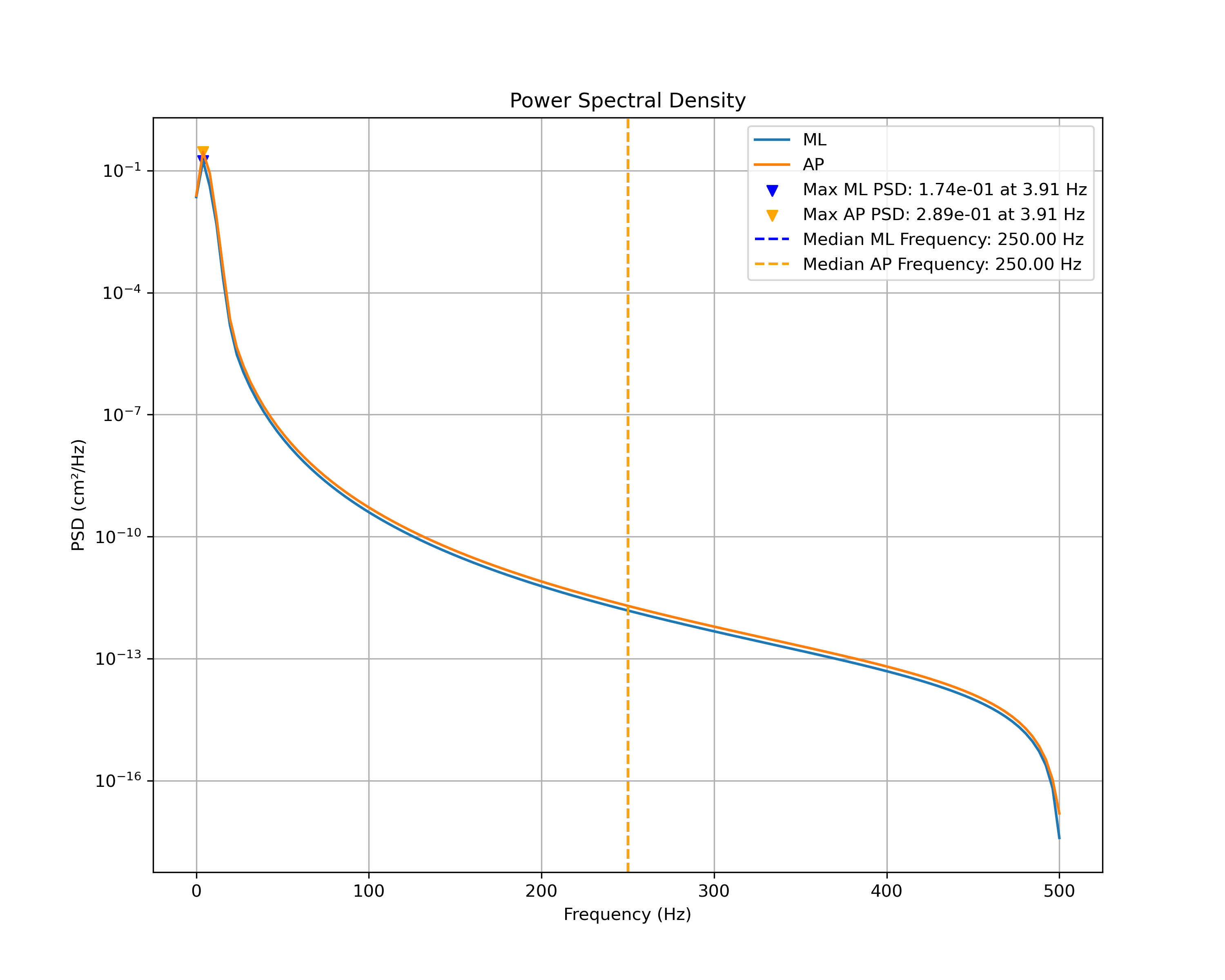}
        \subcaption{Power spectral density (PSD).}\label{fig:cop_psd}
    \end{minipage}
    
    \vspace{0.3cm} 

    \begin{minipage}[t]{0.45\textwidth}
        \centering
        \includegraphics[width=0.95\textwidth]{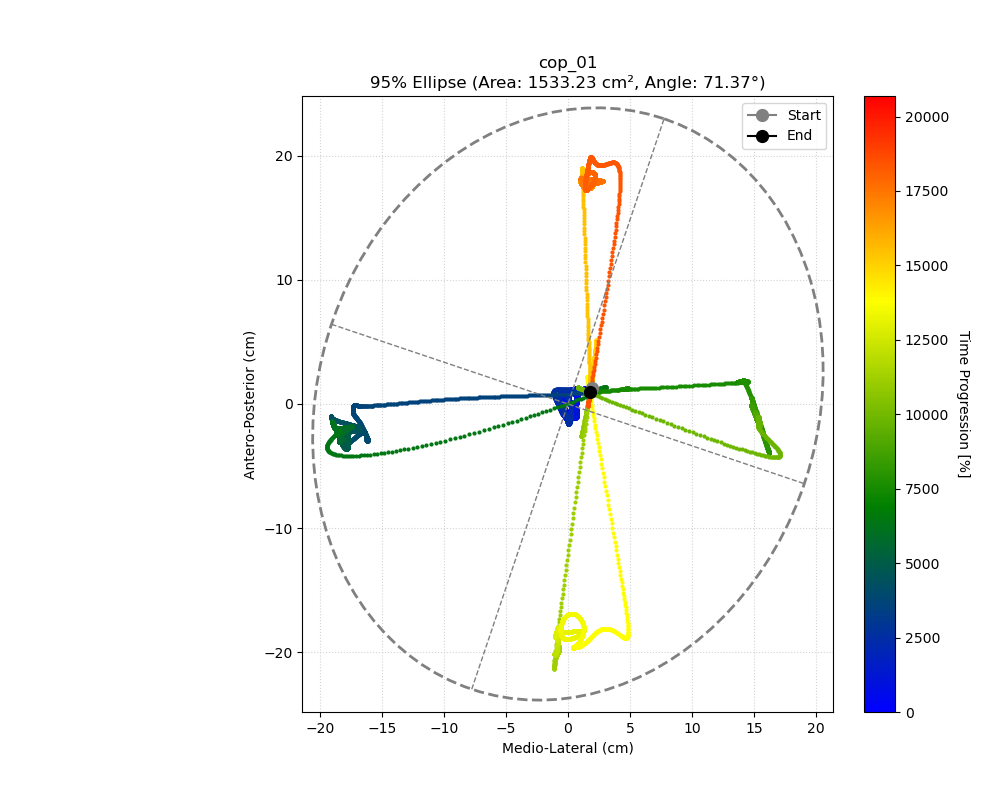}
        \subcaption{General overview of CoP analysis.}\label{fig:cop_overview}
    \end{minipage}
    \hfill
    \begin{minipage}[t]{0.45\textwidth}
        \centering
        \includegraphics[width=0.95\textwidth]{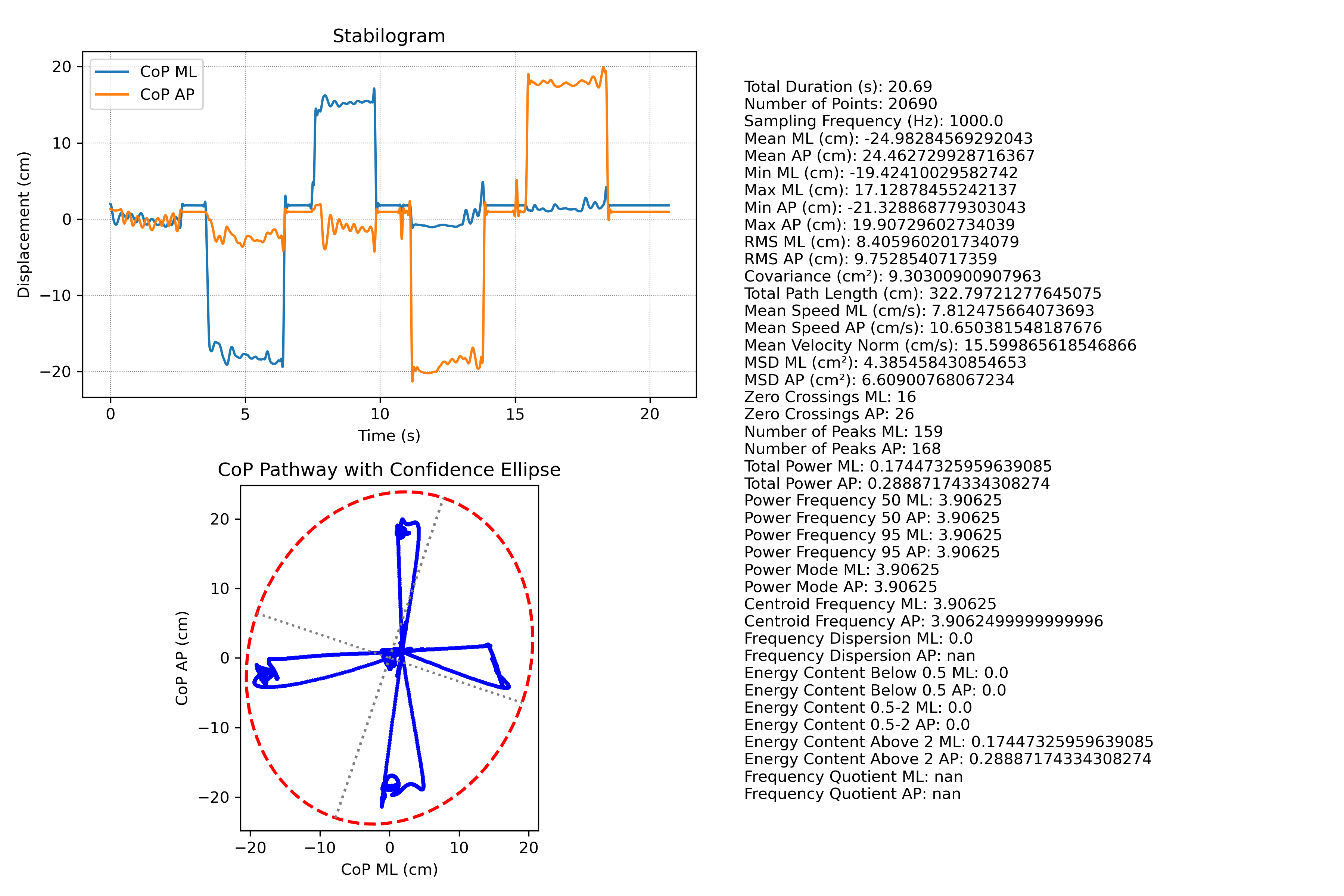}
        \subcaption{Stabilogram, CoP and outputs.}\label{fig:cop_final}
    \end{minipage}
    \caption{Center of Pressure (CoP) analysis outputs: stabilogram (a), PSD (b), general overview (c), and final Stabilogram and CoP path with ellipse 95\% figure (d).}
    \label{fig:cop_analysis}
\end{figure}

Figure \ref{fig:cop_analysis} shows an example of the CoP analysis outputs: the stabilogram, PSD, general CoP overview, and the final figure showing the CoP path over time.

\section{Discussion}

\textit{vailá} represents a significant advancement in biomechanical data analysis by offering a comprehensive, multimodal, open-source toolbox that integrates various data sources into a single environment. Its development addresses the complexities and time-consuming nature of processing kinematic data, particularly from video sources, as highlighted in recent studies~\citep{palucci2022automatic,bini2023criterion,dos2024accuracy,monteiro2024enhancing}. For instance, Palucci Vieira et al.\cite{palucci2022automatic} demonstrated the substantial time investment required for 3D movement kinematic analysis in soccer using markerless motion detection methods compared to traditional digitization. Similarly, Bini et al.\citep{bini2023criterion} discussed the challenges in assessing lower limb motion during cycling with neural networks, emphasizing the need for specialized algorithms and computational resources.

In addition to kinematic data, the integration of kinetic data is crucial for a comprehensive understanding of human movement biomechanics. Kinetic analyses provide insights into the forces and moments acting on the body, which are essential for evaluating balance, stability, and functional performance. Berg-Poppe et al.~\citep{berg2018concurrent} examined the concurrent validity between a portable force plate and an instrumented walkway when measuring limits of stability, highlighting the importance of accurate kinetic measurements in identifying balance deficits that can underlie various disorders. Their findings emphasize the need for accessible tools capable of processing both kinematic and kinetic data to fully assess human movement.

These studies underscore the necessity for efficient, accessible, and user-friendly tools in biomechanics research. Free and open-source tools facilitate research by reducing costs and have the potential for greater citation impact due to their accessibility and adaptability~\citep{vieira2017tracking}. For example, Vieira et al.~\citep{vieira2017tracking} developed a method for tracking futsal players using a wide-angle lens camera, providing their code freely at \href{https://webctim.ulpgc.es/demo112/}{link} and \href{https://github.com/paulopreto/Radial-Distortion-Correction-Sports-Video-Sequences}{GitHub}. Such availability promotes transparency and allows other researchers to build upon their work.

Several free tools like Tracker~\citep{wee2013open}, Kinovea~\citep{abd2015reliability}, and Dvideow~\citep{figueroa2003flexible} have been instrumental in advancing biomechanical analyses by providing accessible platforms for motion tracking and analysis. Tracker, for instance, is widely used for physics education and research due to its simplicity and effectiveness in tracking motion from videos. Kinovea offers user-friendly video analysis features suitable for sports and medical applications. Dvideow~\citep{figueroa2003flexible} provides a flexible software solution for tracking markers in human motion analysis.

While these tools have significantly contributed to the field, \textit{vailá} has the potential to advance beyond them by leveraging modern programming languages and technologies. Developed in Python 3.11.9, which includes CPython acceleration, \textit{vailá} benefits from improved performance and the extensive libraries available in the Python ecosystem. Unlike tools developed in C or C++, which may deter users due to their complexity and less widespread use among the general research community, Python's popularity and readability make \textit{vailá} more accessible to a broader audience. The large and active Python community further strengthens the development and adoption of \textit{vailá}, as users can easily find support, resources, and collaborators.

Moreover, \textit{vailá} shares similarities with tools like DeepLabCut~\citep{nath2019using}, which package scripts in a user-friendly manner to facilitate markerless pose estimation across species and behaviors. Like DeepLabCut, \textit{vailá} leverages open-source development to make advanced computational tools accessible to a broader audience. However, \textit{vailá} extends this approach by integrating multiple modalities of biomechanical data, including kinetic data, offering a more holistic analysis platform.

A distinctive feature of \textit{vailá} is its commitment to open-source principles and transparency. Unlike proprietary software that often uses executable installers (\texttt{.exe}), which can obscure the underlying code and discourage user modification, \textit{vailá} keeps its codebase transparent and accessible on GitHub~\href{https://github.com/vaila-multimodaltoolbox/vaila}{vaila-multimodaltoolbox}. This openness fosters community contributions and encourages users to engage with the code, potentially inspiring them to develop their programming skills in Python. By avoiding closed installers, \textit{vailá} ensures that users can see and modify the code, promoting a deeper understanding of the software's functionality and facilitating customization to meet specific research needs.

The ease of installation and use of \textit{vailá} has been validated by practitioners in laboratory settings. The comprehensive documentation~\href{https://vaila.readthedocs.io/en/latest/}{\textit{vailá} in Read the Docs} guides users through the installation process and demonstrates how to leverage the tool's capabilities effectively. Originally developed to support a specific research project on classifying single-limb motor development using soccer kicking in young children with cerebral palsy, funded by the CAPES Foundation Brazil (CAPES process: 88887.936639/2024-00), \textit{vailá} has since grown into a versatile tool ready for broader application and community contribution.

Despite its current capabilities, \textit{vailá} acknowledges its limitations and areas for future development. While it offers functionalities such as the Markerless 2D Workflow, Motion Capture Cluster Workflow, EMG Workflow, and Center of Pressure (CoP) Balance Analysis, there is ample room for expansion. Future perspectives include enhancing the user interface for even greater intuitiveness, incorporating advanced machine learning algorithms for data analysis, and extending support to additional data types and devices. Continuous development and community engagement are essential to realize these enhancements.

Furthermore, in this article, we have presented examples of workflows and pipelines within \textit{vailá} that facilitate rapid processing of large datasets in batch mode, independent of the type of data collection. This flexibility is particularly valuable in research scenarios where unexpected opportunities arise and there is a need to process data quickly without losing valuable collections. By enabling swift adaptation to varying data types and collection methods, \textit{vailá} helps researchers efficiently manage and analyze data, even under time constraints or unforeseen circumstances.

The creation of \textit{vailá} was born out of a practical need in a laboratory environment where researchers required modified and versatile code daily for each data collection. This necessity mirrors the sentiment expressed in Albert Einstein's often-misunderstood quote, "Imagination is more important than knowledge." Contrary to common misinterpretation, Einstein did not downplay the importance of knowledge but emphasized that imagination is the driving force behind innovation and the application of knowledge in new ways. In the context of \textit{vailá}, imagination has been the catalyst for developing a tool that challenges traditional norms, offering a flexible and open platform that empowers researchers to explore, experiment, and innovate without constraints.

By embracing imagination and creativity, \textit{vailá} has evolved beyond its initial scope, providing a foundation for researchers and laboratories eager to experiment with new computational approaches. Its design encourages users not only to utilize existing functionalities but also to contribute new modules and workflows, fostering a collaborative environment where knowledge and imagination synergize to advance the field of biomechanics.

\textit{vailá} offers a flexible, accessible, and comprehensive solution to the challenges of biomechanical data analysis. By simplifying data processing, integrating multiple data sources—including kinetic data as highlighted by Berg-Poppe et al.~\citep{berg2018concurrent}—and promoting open-source development, \textit{vailá} has the potential to empower researchers and practitioners to conduct more efficient and inclusive biomechanical studies. Its contributions could significantly advance the field, leading to improved rehabilitation strategies, performance optimization, and a deeper understanding of human movement. As the tool develops, it invites the community to participate actively in its evolution, embodying the spirit of collaborative innovation essential for scientific progress.

\section{Conclusion}

\textit{vailá} offers a flexible, accessible, and comprehensive solution to the challenges of biomechanical data analysis. By simplifying data processing, integrating multiple data sources, and promoting open-source development, \textit{vailá} empowers researchers and practitioners to conduct more efficient and inclusive biomechanical studies. Its contributions have the potential to significantly advance the field, leading to improved rehabilitation strategies, performance optimization, and a deeper understanding of human movement. As the tool develops, it invites the community to participate actively in its evolution, embodying the spirit of collaborative innovation essential for scientific progress.
\section*{Acknowledgements}

We would like to express our gratitude to the School of Physical Education and Sports of Ribeirão Preto, University of São Paulo, and to the Ribeirão Preto Medical School, University of São Paulo, Ribeirão Preto, Brazil, for providing salary support and a leave of absence during the research period. We would also like to thank the University of North Florida (UNF) for hosting the six-month research internship abroad.

\section*{Disclosure statement}

The authors declare that they have no competing financial interests or personal relationships that could have appeared to influence the work reported in this paper. The project is open-source, entirely free, and non-commercial, aimed solely at those interested in studying and advancing biomechanical analysis.

\section*{Funding}

This study was financed in part by the Coordenação de Aperfeiçoamento de Pessoal de Nível Superior – Brasil (CAPES) – Finance Code 001; by PRINT-USP-CAPES Foundation for supporting the Senior Visiting Professor program at the University of North Florida under process number 88887.936639/2024-00; and by FAPESP under the grant number \#2019/17729-0. This project  was also partially supported financially by UNF MedNexus Research Innovation Fund (Cesar, GM), and the Dean’s Office for Research and Innovation of the University of São Paulo (Support Program for New Professors - Bedo, BLS).

\section*{Availability of Data and Materials}
\sloppy
The \textit{vailá} toolbox is available at our GitHub repository: \url{https://github.com/vaila-multimodaltoolbox/vaila}. All code is open-source, and we encourage contributions from researchers, developers, and users. The repository includes detailed documentation, example datasets, and installation scripts for Linux, macOS, and Windows.

We welcome feedback, suggestions, and code contributions to help improve and expand \textit{vailá} to meet the needs of the biomechanics community. Please feel free to open issues, submit pull requests, or join discussions to contribute to the project.


\bibliographystyle{unsrtnat}
\bibliography{references}  

\begin{thebibliography}{27}
\providecommand{\natexlab}[1]{#1}
\providecommand{\url}[1]{\texttt{#1}}
\expandafter\ifx\csname urlstyle\endcsname\relax
  \providecommand{\doi}[1]{doi: #1}\else
  \providecommand{\doi}{doi: \begingroup \urlstyle{rm}\Url}\fi

\bibitem[Ricamato and Hidler(2005)]{ricamato2005quantification}
Anthony~L Ricamato and Joseph~M Hidler.
\newblock Quantification of the dynamic properties of emg patterns during gait.
\newblock \emph{Journal of electromyography and kinesiology}, 15\penalty0
  (4):\penalty0 384--392, 2005.

\bibitem[Roberts et~al.(2017)Roberts, Mongeon, and Prince]{roberts2017}
M.~Roberts, D.~Mongeon, and F.~Prince.
\newblock Biomechanical parameters for gait analysis: a systematic review of
  healthy human gait.
\newblock \emph{Phys. Ther. Rehabil}, 4\penalty0 (6), 2017.
\newblock \doi{10.7243/2055-2386-4-6}.

\bibitem[Cesar et~al.(2016)Cesar, Tomasevicz, and Burnfield]{cesar2016frontal}
Guilherme~M Cesar, Curtis~L Tomasevicz, and Judith~M Burnfield.
\newblock Frontal plane comparison between drop jump and vertical jump:
  implications for the assessment of acl risk of injury.
\newblock \emph{Sports Biomechanics}, 15\penalty0 (4):\penalty0 440--449, 2016.

\bibitem[Bedo et~al.(2021)Bedo, Cesar, Moraes, Mariano, Vieira, Andrade, and
  Santiago]{bedo2021influence}
Bruno Luiz~Souza Bedo, Guilherme~Manna Cesar, Renato Moraes, F{\'a}bio~Pamplona
  Mariano, Luiz Henrique~Palucci Vieira, Vitor~Luiz Andrade, and Paulo
  Roberto~Pereira Santiago.
\newblock Influence of side uncertainty on knee kinematics of female handball
  athletes during sidestep cutting maneuvers.
\newblock \emph{Journal of applied biomechanics}, 37\penalty0 (3):\penalty0
  188--195, 2021.

\bibitem[Cesar et~al.(2022)Cesar, Buster, Gonabadi, and
  Burnfield]{cesar2022muscle}
Guilherme~M Cesar, Thad~W Buster, Arash~Mohammadzadeh Gonabadi, and Judith~M
  Burnfield.
\newblock Muscle demand and kinematic similarities between pediatric-modified
  motor-assisted elliptical training at fast speed and fast overground walking:
  Real-world implications for pediatric gait rehabilitation.
\newblock \emph{Journal of electromyography and kinesiology}, 63:\penalty0
  102639, 2022.

\bibitem[Barre and Armand(2014)]{Barre2014}
A.~Barre and S.~Armand.
\newblock Biomechanical toolkit: Open-source framework to visualize and process
  biomechanical data.
\newblock \emph{Comput Methods Programs Biomed}, 114\penalty0 (1):\penalty0
  80--87, 2014.

\bibitem[Seth et~al.(2018)Seth, Hicks, Uchida, Habib, Dembia, Dunne, Ong,
  DeMers, Rajagopal, Millard, et~al.]{seth2018opensim}
Ajay Seth, Jennifer~L Hicks, Thomas~K Uchida, Ayman Habib, Christopher~L
  Dembia, James~J Dunne, Carmichael~F Ong, Matthew~S DeMers, Apoorva Rajagopal,
  Matthew Millard, et~al.
\newblock Opensim: Simulating musculoskeletal dynamics and neuromuscular
  control to study human and animal movement.
\newblock \emph{PLoS computational biology}, 14\penalty0 (7):\penalty0
  e1006223, 2018.

\bibitem[Bedo et~al.(2020{\natexlab{a}})Bedo, Catelli, Lamontagne, and
  Santiago]{bedo2020custom}
Bruno L.~S. Bedo, Danilo~S. Catelli, Mario Lamontagne, and Paulo R.~P.
  Santiago.
\newblock A custom musculoskeletal model for estimation of medial and lateral
  tibiofemoral contact forces during tasks with high knee and hip flexions.
\newblock \emph{Computer Methods in Biomechanics and Biomedical Engineering},
  23\penalty0 (10):\penalty0 658--663, 2020{\natexlab{a}}.

\bibitem[Bedo et~al.(2020{\natexlab{b}})Bedo, Mantoan, Catelli, Cruaud,
  Reggiani, and Lamontagne]{bedo2020474}
Bruno L.~S. Bedo, A~Mantoan, D.~S. Catelli, W.~Cruaud, M.~Reggiani, and
  M.~Lamontagne.
\newblock 474 bops: a matlab toolbox to batch musculoskeletal data processing
  for opensim.
\newblock \emph{Computer 475 Methods in Biomechanics and Biomedical
  Engineering}, pages 1--11, 2020{\natexlab{b}}.

\bibitem[van Rossum and the Python Development~Team(2022)]{python311}
Guido van Rossum and the Python Development~Team.
\newblock Python 3.11.0 release.
\newblock
  \url{https://docs.python.org/3/whatsnew/3.11.html\#whatsnew311-faster-cpython},
  2022.
\newblock Accessed: 2024-09-24.

\bibitem[Smith et~al.(2023)Smith, Thompson, and et~al.]{Smith2023}
J.~M. Smith, A.~K. Thompson, and et~al.
\newblock Markerless motion capture using deep learning techniques: A
  systematic review.
\newblock \emph{Computer Methods in Biomechanics and Biomedical Engineering},
  2023.
\newblock \doi{10.1016/j.compbiomed.2023.104210}.

\bibitem[Lugaresi et~al.(2019)Lugaresi, Tang, Nash, McClanahan, Uboweja, Hays,
  Zhang, Chang, Yong, Lee, Chang, Hua, Georg, and
  Grundmann]{lugaresi2019mediapipe}
Camillo Lugaresi, Jiuqiang Tang, Hadon Nash, Chris McClanahan, Esha Uboweja,
  Michael Hays, Fan Zhang, Chuo-Ling Chang, Ming~Guang Yong, Juhyun Lee,
  Wan-Teh Chang, Wei Hua, Manfred Georg, and Matthias Grundmann.
\newblock Mediapipe: A framework for building perception pipelines, 2019.
\newblock URL \url{https://arxiv.org/abs/1906.08172}.

\bibitem[Matthis et~al.(2022)Matthis, Cherian, and Wirth]{Matthis2022}
Jonathan Matthis, Aaron Cherian, and Trent Wirth.
\newblock The freemocap project - and - gaze/hand coupling during a combined
  three-ball juggling and balance task.
\newblock \emph{Journal of Vision}, 22\penalty0 (14):\penalty0 4195, 2022.
\newblock \doi{10.1167/jov.22.14.4195}.

\bibitem[Wee(2013)]{wee2013open}
Loo~Kang Wee.
\newblock Open source physics.
\newblock \emph{arXiv preprint arXiv:1308.2614}, 2013.

\bibitem[Figueroa et~al.(2003)Figueroa, Leite, and
  Barros]{figueroa2003flexible}
Pascual~J Figueroa, Neucimar~J Leite, and Ricardo~ML Barros.
\newblock A flexible software for tracking of markers used in human motion
  analysis.
\newblock \emph{Computer methods and programs in biomedicine}, 72\penalty0
  (2):\penalty0 155--165, 2003.

\bibitem[Brooks~Jr(1995)]{brooks1995mythical}
Frederick~P Brooks~Jr.
\newblock \emph{The mythical man-month: essays on software engineering}.
\newblock Pearson Education, 1995.

\bibitem[Brooks(1995)]{brooks1995mythical_20years}
Frederick~P. Brooks.
\newblock The mythical man-month: After 20 years.
\newblock \emph{IEEE Software}, 12\penalty0 (5):\penalty0 57--60, 1995.
\newblock \doi{10.1109/MS.1995.10042}.

\bibitem[GPL(2007)]{GPLv3}
Gnu general public license, version 3.
\newblock \url{http://www.gnu.org/licenses/gpl.html}, June 2007.
\newblock Last retrieved 2020-01-01.

\bibitem[Duarte and Freitas(2010)]{duarte2010revision}
Marcos Duarte and Sandra~MSF Freitas.
\newblock Revision of posturography based on force plate for balance
  evaluation.
\newblock \emph{Brazilian Journal of physical therapy}, 14:\penalty0 183--192,
  2010.

\bibitem[Palucci~Vieira et~al.(2022)Palucci~Vieira, Santiago, Pinto, Aquino,
  Torres, and Barbieri]{palucci2022automatic}
Luiz~H Palucci~Vieira, Paulo~RP Santiago, Allan Pinto, Rodrigo Aquino, Ricardo
  da~S Torres, and Fabio~A Barbieri.
\newblock Automatic markerless motion detector method against traditional
  digitisation for 3-dimensional movement kinematic analysis of ball kicking in
  soccer field context.
\newblock \emph{International journal of environmental research and public
  health}, 19\penalty0 (3):\penalty0 1179, 2022.

\bibitem[Bini et~al.(2023)Bini, Serrancoli, Santiago, Pinto, and
  Moura]{bini2023criterion}
Rodrigo~Rico Bini, Gil Serrancoli, Paulo Roberto~Pereira Santiago, Allan Pinto,
  and Felipe Moura.
\newblock Criterion validity of neural networks to assess lower limb motion
  during cycling.
\newblock \emph{Journal of sports sciences}, 41\penalty0 (1):\penalty0 36--44,
  2023.

\bibitem[dos Santos~Banks et~al.(2024)dos Santos~Banks, Santiago,
  da~Silva~Torres, de~Oliveira, and Moura]{dos2024accuracy}
Lu{\'\i}za dos Santos~Banks, Paulo Roberto~Pereira Santiago, Ricardo
  da~Silva~Torres, Donizete Cicero~Xavier de~Oliveira, and Felipe~Arruda Moura.
\newblock Accuracy of a markerless system to estimate the position of taekwondo
  athletes in an official combat area.
\newblock \emph{International Journal of Performance Analysis in Sport}, pages
  1--16, 2024.

\bibitem[Monteiro et~al.(2024)Monteiro, Dos~Santos, Blauberger, Link,
  Russomanno, Tahara, Chinaglia, and Santiago]{monteiro2024enhancing}
Rafael Luiz~Martins Monteiro, Carlos Cesar~Arruda Dos~Santos, Patrick
  Blauberger, Daniel Link, Tiago~Guedes Russomanno, Ariany~Klein Tahara,
  Abel~Gon{\c{c}}alves Chinaglia, and Paulo Roberto~Pereira Santiago.
\newblock Enhancing soccer goalkeepers penalty dive kinematics with
  instructional video and laterality insights in field conditions.
\newblock \emph{Scientific Reports}, 14\penalty0 (1):\penalty0 10225, 2024.

\bibitem[Berg-Poppe et~al.(2018)Berg-Poppe, Cesar, Tao, Johnson, and
  Landry]{berg2018concurrent}
Patti Berg-Poppe, Guilherme~M Cesar, Hanz Tao, Cal Johnson, and Jessica Landry.
\newblock Concurrent validity between a portable force plate and instrumented
  walkway when measuring limits of stability.
\newblock \emph{International Journal of Therapy and Rehabilitation},
  25\penalty0 (6):\penalty0 272--278, 2018.

\bibitem[Vieira et~al.(2017)Vieira, Pagnoca, Milioni, Barbieri, Menezes,
  Alvarez, D{\'e}niz, Santana-Cedr{\'e}s, and Santiago]{vieira2017tracking}
Luiz~HP Vieira, Emilio~A Pagnoca, Fabio Milioni, Ricardo~A Barbieri, Rafael~P
  Menezes, Luis Alvarez, Luis~G D{\'e}niz, Daniel Santana-Cedr{\'e}s, and
  Paulo~RP Santiago.
\newblock Tracking futsal players with a wide-angle lens camera: accuracy
  analysis of the radial distortion correction based on an improved hough
  transform algorithm.
\newblock \emph{Computer Methods in Biomechanics and Biomedical Engineering:
  Imaging \& Visualization}, 5\penalty0 (3):\penalty0 221--231, 2017.

\bibitem[Abd El-Raheem et~al.(2015)Abd El-Raheem, Kamel, and
  Ali]{abd2015reliability}
Reham~M Abd El-Raheem, Ragia~M Kamel, and Mohammad~F Ali.
\newblock Reliability of using kinovea program in measuring dominant wrist
  joint range of motion.
\newblock \emph{Trends in Applied Sciences Research}, 10\penalty0 (4):\penalty0
  224, 2015.

\bibitem[Nath et~al.(2019)Nath, Mathis, Chen, Patel, Bethge, and
  Mathis]{nath2019using}
Tanmay Nath, Alexander Mathis, An~Chi Chen, Amir Patel, Matthias Bethge, and
  Mackenzie~Weygandt Mathis.
\newblock Using deeplabcut for 3d markerless pose estimation across species and
  behaviors.
\newblock \emph{Nature protocols}, 14\penalty0 (7):\penalty0 2152--2176, 2019.

\end{thebibliography}
\par

\end{document}